\documentclass{IEEEojcsys}

\usepackage[colorlinks,urlcolor=blue,linkcolor=blue,citecolor=blue]{hyperref}

\let\labelindent\relax
\usepackage{enumitem}

\usepackage{cite}

\usepackage{dsfont}
\usepackage{amsfonts}
\usepackage{textcomp}

\usepackage{amsmath}
\usepackage{amssymb}
\usepackage{commath}
\usepackage{derivative}

\usepackage{graphicx}
\usepackage{adjustbox}
\usepackage{caption}
\usepackage{svg}

\usepackage{pgfplots}
\usepackage{tikz}

\pgfplotsset{compat=newest}
\usetikzlibrary{decorations.pathmorphing}
\usetikzlibrary{fit}
\usetikzlibrary{backgrounds}
\usetikzlibrary{pgfplots.groupplots}

\usepackage{url}
\usepackage{booktabs}
\usepackage{siunitx}
\usepackage{xcolor}

\usepackage[acronym, shortcuts]{glossaries}

\usepackage{hyperref}
\hypersetup{
	colorlinks=true,
	linkcolor=blue,
	filecolor=magenta,      
	urlcolor=blue,
}

\usepackage[ruled]{algorithm2e}
\makeatletter
\renewcommand{\algocf@caption@boxruled}{%
	\hrule
	\hbox to \hsize{%
		\vrule\hskip-0.6pt
		\vbox{   
			\vskip\interspacetitleboxruled%
			\vspace{0.1em}
			\unhbox\algocf@capbox\hfill
			\vspace{0.1em}
			\vskip\interspacetitleboxruled
		}%
		\hskip-0.6pt\vrule%
	}\nointerlineskip%
}%

\makeatletter
\newcommand{\removealgorithmerror}{\let\@latex@error\@gobble}
\makeatother

\usepackage{bm}

\renewcommand{\vec}[1]{\bm{\mathbf{#1}}}
\newcommand{\mat}[1]{\bm{\mathbf{#1}}}

\newcommand{\hvec}[1]{\hat{\bm{\mathbf{#1}}}}
\newcommand{\hmat}[1]{\hat{\bm{\mathbf{#1}}}}

\DeclareMathOperator*{\argmax}{arg\,max}
\DeclareMathOperator*{\argmin}{arg\,min}
\DeclareMathOperator*{\st}{s.t.}

\DeclareMathOperator{\kl}{KL}

\newcommand*\p{\textnormal{\textquotesingle}}

\DeclareOldFontCommand{\bf}{\normalfont\bfseries}{\mathbf}

\newcommand{\op}[1]{\operatorname{#1}}

\newcommand{\Dir}{\op{Dir}}
\newcommand{\Cat}{\op{Cat}}

\newcommand{\N}{\op{N}}
\newcommand{\NW}{\op{NW}}

\newcommand{\MNW}{\op{MNW}}

\newcommand{\W}{\op{W}}

\newacronym{EM}{EM}{expectation-maximization}
\newacronym{RL}{RL}{reinforcement learning}
\newacronym{Hb-REPS}{Hb-REPS}{hybrid relative entropy policy search}
\newacronym{SDS}{SDS}{switching dynamical systems}
\newacronym{SSM}{SSM}{switching state-space models}
\newacronym{PWL}{PWL}{piecewise linear}
\newacronym{ReLU}{ReLU}{rectified linear unit}
\newacronym{PWA}{PWA}{piecewise affine}
\newacronym{PWARX}{PWARX}{piecewise autoregressive exogenous systems}
\newacronym{MIQP}{MIQP}{mixed-integer quadratic program}
\newacronym{PGM}{PGM}{probabilistic graphical models}
\newacronym{HDBN}{HDBN}{hybrid dynamic Bayesian networks}
\newacronym{BNP}{BNP}{Bayesian nonparametric}
\newacronym{MCMC}{MCMC}{Markov chain Monte Carlo}
\newacronym{SVI}{SVI}{stochastic variational inference}
\newacronym{HRL}{HRL}{hierarchical reinforcement learning}
\newacronym{SMDP}{SMDP}{semi-Markov decision processes}
\newacronym{MDP}{MDP}{Markov decision processes}
\newacronym{rARHMM}{rARHMM}{recurrent autoregressive hidden Markov model}
\newacronym{ARHMM}{ARHMM}{autoregressive hidden Markov model}
\newacronym{HMM}{HMM}{hidden Markov model}
\newacronym{HSMM}{HSMM}{semi-hidden Markov model}
\newacronym{SLDS}{SLDS}{switching linear dynamical systems}
\newacronym{rSLDS}{rSLDS}{recurrent switching linear dynamical systems}
\newacronym{FHMM}{FHMM}{factorial hidden Markov model}
\newacronym{REPS}{REPS}{relative entropy policy search}
\newacronym{KL}{KL}{Kullback-Leibler divergence}
\newacronym{MAP}{MAP}{maximum a posteriori}
\newacronym{NW}{NW}{normal-Wishart}
\newacronym{MNW}{MNW}{matrix-normal-Wishart}
\newacronym{Dir}{Dir}{Dirichlet}
\newacronym{Cat}{Cat}{categorical}
\newacronym{E-step}{E-step}{expectation step}
\newacronym{M-step}{M-step}{maximization step}
\newacronym{EB-step}{EB-step}{empirical Bayes step}
\newacronym{FNN}{FNN}{feed-forward neural net}
\newacronym{GP}{GP}{Gaussian process}
\newacronym{LSTM}{LSTM}{long-short-term memory network}
\newacronym{RNN}{RNN}{recurrent neural network}
\newacronym{NMSE}{NMSE}{normalized mean square error}
\newacronym{SAC}{SAC}{soft actor-critic}
\newacronym{RFF}{RFF}{random Fourier feature}
\newacronym{TD}{TD}{temporal difference}
\newacronym{MPC}{MPC}{model predictive control}

\jvol{00}
\jnum{XX}
\paper{1234567}
\pubyear{2021}
\receiveddate{XX September 2021}
\accepteddate{XX October 2021}
\publisheddate{XX November 2021}
\currentdate{XX November 2021}
\doiinfo{OJCSYS.2021.Doi Number}

\begin{document}
	
\sptitle{Article Category}
\title{Model-Based Reinforcement Learning \\ via Stochastic Hybrid Models}
\editor{This paper was recommended by Associate Editor F. A. Author.}
\author{Hany Abdulsamad \affilmark{1}}
\author{Jan Peters \affilmark{2} (Fellow, IEEE)}
\affil{Department of Electrical Engineering and Automation, Aalto University, Finland.} 
\affil{Department of Computer Science, Technical University of Darmstadt, Germany.} 
\corresp{CORRESPONDING AUTHOR: Hany Abdulsamad (e-mail: \href{mailto:hany.abdulsamad@aalto.fi}{hany.abdulsamad@aalto.fi})}

\markboth{PREPARATION OF PAPERS FOR IEEE OPEN JOURNAL OF CONTROL SYSTEMS}{F. A. AUTHOR {\itshape ET AL}.}

\begin{abstract}
	Optimal control of general nonlinear systems is a central challenge in automation. Enabled by powerful function approximators, data-driven approaches to control have recently successfully tackled challenging applications. However, such methods often obscure the structure of dynamics and control behind black-box over-parameterized representations, thus limiting our ability to understand closed-loop behavior. This paper adopts a hybrid-system view of nonlinear modeling and control that lends an explicit hierarchical structure to the problem and breaks down complex dynamics into simpler localized units. We consider a sequence modeling paradigm that captures the temporal structure of the data and derive an \gls*{EM} algorithm that automatically decomposes nonlinear dynamics into stochastic piecewise affine models with nonlinear transition boundaries. Furthermore, we show that these time-series models naturally admit a closed-loop extension that we use to extract local polynomial feedback controllers from nonlinear experts via behavioral cloning. Finally, we introduce a novel \gls*{Hb-REPS} technique that incorporates the hierarchical nature of hybrid models and optimizes a set of time-invariant piecewise feedback controllers derived from a piecewise polynomial approximation of a global state-value function.
\end{abstract}

\begin{IEEEkeywords}
	Reinforcement Learning, Hybrid Models, Bayesian Inference, Hidden Markov Models, Expectation-Maximization, System Identification, Behavioral Cloning, Piecewise Feedback Control.
\end{IEEEkeywords}

\maketitle

\section{Introduction}
\label{sec:intro}
The class of nonlinear dynamical systems governs a vast range of real-world applications and underpins the most challenging problems in classical control, and \gls*{RL} \cite{fantoni2002non, kober2013reinforcement}. Recent developments in learning-for-control have pushed towards deploying more complex and highly sophisticated representations, e.g., (deep) neural networks and Gaussian processes, to capture the structure of both dynamics and controllers. This trend led to unprecedented success in the domain of \gls*{RL} \cite{mnih2015human} and can be observed in both approximate optimal control \cite{deisenroth2011pilco, levine2016end, hafner2019learning}, and approximate value and policy iteration algorithms \cite{schulman2015trust, lillicrap2015continuous, haarnoja2018soft}.

However, before the latest revival of neural networks, research has focused on different paradigms for solving complex control tasks. One interesting concept relied on decomposing nonlinear structures of dynamics and control into simpler piecewise (affine) components, each responsible for an area of the state-action space. 
Instances of this abstraction can be found in the control literature under the labels of hybrid systems or switched models \cite{liberzon2003switching, haddad2006impulsive, goebel2012hybrid, borrelli2017predictive}, while in the machine and reinforcement learning communities, the terminology of switching dynamical systems and hybrid state-space models is more widely used \cite{ghahramani2000variational, beal2003variational, fox2009bayesian, linderman2017recurrent}. 

While the hybrid-state paradigm is a natural choice for studying jump processes, it also provides a surrogate piecewise approximation of general nonlinear dynamical behavior. Despite being less flexible than generic black-box approximators, hybrid models can regularize functional complexity and contribute to improved interpretability by imposing a structured representation.   

Adopting this perspective in this paper, we present techniques for data-driven automatic system identification and closed-loop control of general nonlinear systems using piecewise polynomial hybrid surrogate models. More concretely, we focus on dynamic Bayesian graphical models as hybrid representations due to their favorable properties. These models have an inherent time-recurrent structure that captures correlations over extended horizons and carry over the advantages of well-established recursive Bayesian inference techniques for dynamical time series data.

In prior work \cite{abdulsamad2020hierarchical}, we presented a maximum likelihood approach for hierarchical piecewise system identification and behavioral cloning. Here, we robustify that approach by introducing suitable priors over all parameters. However, the central contribution of this paper is the introduction of an infinite horizon reinforcement learning framework that integrates the structured representation of stochastic hybrid models. The resulting algorithm interactively synthesizes nonlinear feedback controllers and value functions via a hierarchical piecewise polynomial architecture. 

This paper is structured as follows. In Section~\ref{sec:related}, we start by reviewing and comparing prominent paradigms of system modeling and optimal control of hybrid systems. Using that context in Section~\ref{sec:contrib}, we highlight the advantages of our contributions in comparison with the literature. In Section~\ref{sec:problem}, we cast the control problem as an infinite horizon Markov decision process and extended it to accommodate a hybrid structure. Next, in Section~\ref{sec:hdbn}, we introduce our notation of stochastic switching models in the form of hybrid dynamic Bayesian networks, as previously established in \cite{abdulsamad2020hierarchical}. In Section~\ref{sec:sysid}, we recap our approach from \cite{abdulsamad2020hierarchical} and improve it to derive a maximum a posteriori \acrfull*{EM} algorithm for inferring the parameters of probabilistic hybrid models from data. This inference method is helpful for automatically decomposing nonlinear open-loop dynamics into switching affine regimes with arbitrary boundaries and deconstructing state-of-the-art nonlinear expert controllers into piecewise polynomial policies. Furthermore, in Section~\ref{sec:rl}, we formulate hybrid optimal control as a stochastic optimization problem and derive a trust-region reinforcement learning algorithm that incorporates an explicit hierarchical model of the nonlinear dynamics. We use this approach to iteratively learn piecewise approximations of the global nonlinear value function and stationary feedback controller. Finally, in Section~\ref{sec:eval}, we empirically evaluate our approaches on examples of stochastic nonlinear systems, including results from \cite{abdulsamad2020hierarchical} that contribute to the overall picture.

Our empirical evaluation indicates that hybrid models can provide an alternative to generic black-box representations for system identification, behavioral cloning, and learning-based control. Hybrid models are able to reach comparable performance and deliver simpler, easily identifiable switching patterns of dynamics and control while requiring a fraction of the number of parameters of other functional forms. However, the results also reveal certain drawbacks, mainly in poor scalability and increased algorithmic complexity. We address these issues in a final outlook in Section~\ref{sec:conclusion}.

\section{Related Work}
\label{sec:related}
This section reviews work related to the modeling and control of hybrid systems and highlights connections and parallels between approaches stemming from the control and machine and reinforcement learning literature.

Hybrid systems have been extensively studied in the control community and are widely used in real-world applications \cite{borrelli2006mpc, menchinelli2008hybrid}. For research on hybrid system identification, we refer to survey work in \cite{paoletti2007identification} and \cite{garulli2012survey}. There, the authors focus on \gls*{PWA} systems and introduce taxonomies of different representations and procedures commonly used for identifying sub-regimes of dynamics, ranging from algebraic approaches \cite{vidal2003algebraic} to mixed-integer optimization \cite{bemporad2001identification}, and Bayesian methods \cite{juloski2005bayesian}. Furthermore, identification techniques for piecewise nonlinear systems have been developed based on sparse optimization \cite{bako2010l} and kernel methods \cite{lauer2010nonlinear}. Finally, it is worth noting that the majority of literature considers deterministic regime-switching events with exceptions in \cite{bemporad2005optimal, cassandras2006stochastic}.

\begin{figure}[!t]
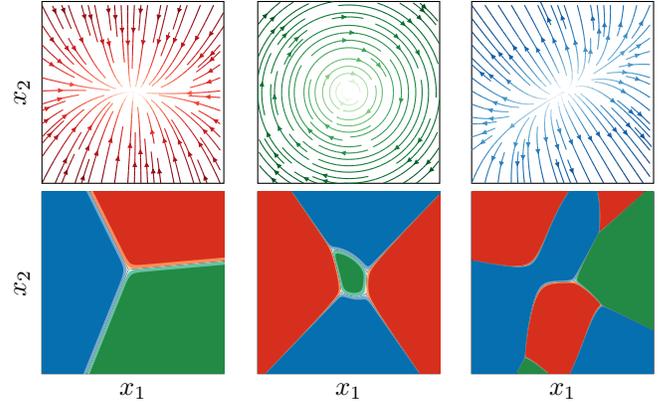

	\begin{minipage}[t]{0.3\columnwidth}%
		\input{figures/hybrid_dyn_0.tex}%
	\end{minipage}\hspace*{.75cm}%
	\begin{minipage}[t]{0.3\columnwidth}%
		\input{figures/hybrid_dyn_1.tex} %
	\end{minipage}\hspace*{.25cm}%
	\begin{minipage}[t]{0.3\columnwidth}%
		\input{figures/hybrid_dyn_2.tex}%
	\end{minipage}%
	\vspace{0.1cm}
	\begin{minipage}[t]{0.3\columnwidth}%
		\input{figures/hybrid_lgstc_linear.tex}%
	\end{minipage}\hspace*{.75cm}%
	\begin{minipage}[t]{0.3\columnwidth}%
		\input{figures/hybrid_lgstc_quad.tex}%
	\end{minipage}\hspace*{.25cm}%
	\begin{minipage}[t]{0.3\columnwidth}%
		\input{figures/hybrid_lgstc_cube.tex}%
	\end{minipage}
	\caption{A hybrid system with $K=3$ piecewise affine regimes. The top row depicts the mean unforced continuous transition dynamics in the phase space. The bottom row shows the distinct activation regions of the three dynamics regime across the phase space. We illustrate examples of affine (left), quadratic (middle), and third-order polynomial (right) switching boundaries. Figure reproduced from \cite{abdulsamad2020hierarchical}.}
	\label{fig:rarhmm_example}
	\vspace{-0.5cm}
\end{figure}

Research in the area of optimal control for hybrid systems stretches back to the seminal work in \cite{sontag1981nonlinear}, which highlights the possibility of general nonlinear control by considering piecewise affine systems. In \cite{zhu2015optimal}, an overview of control approaches for piecewise affine switching dynamics is presented. The authors categorize the literature by distinguishing between externally and internally forced switching mechanisms. The bulk of optimal control approaches in this area focuses on (nonlinear) \gls*{MPC} \cite{camacho2010model}. Here we highlight the influential work in \cite{bemporad1999control}, which formulates the optimal control problem as a \gls*{MIQP}. This approach was later extended in \cite{bemporad2000piecewise} and \cite{borrelli2003efficient} to solve a multi-parametric \gls*{MIQP} and arrive at time-variant piecewise affine state-feedback controllers and piecewise quadratic value functions with polyhedral partitions. Recently, more efficient formulations of hybrid control have been proposed \cite{marcucci2019mixed}, which leverage modern techniques from mixed-integer and disjunctive programming to tackle large-scale problems.

Hybrid representations also play a central role in data-driven, general-purpose process modeling and state estimation \cite{ackerson1970state, hamilton1990analysis}, where different classes of stochastic hybrid systems serve as powerful generative models for complex dynamical behaviors \cite{pavlovic2001learning, oh2005data, mesot2007switching}. The dominant paradigm in this domain has been that of \gls*{PGM}, more specifically, \gls*{HDBN} for temporal modeling \cite{koller2009probabilistic, lerner2002hybrid}. One crucial contribution of recent Bayesian interpretations of switching systems is rooted in the \gls*{BNP} view \cite{escobar1995bayesian, rasmussen1999infinite, beal2002infinite, teh2005sharing}. This perspective theoretically allows for an infinite number of components, thus dramatically increasing the expressiveness of such models. Given the limited scope of this review section, we highlight only recent contributions with high impacts, such as \cite{fox2009nonparametric} and \cite{linderman2017recurrent}, which successfully develop \gls*{MCMC} and \gls*{SVI} techniques for system identification. More recently, the rise of variational auto-encoders \cite{kingma2013auto} has enabled a new and powerful view of inference techniques \cite{becker2019switching} for hybrid systems. A distinct drawback of such approaches is their reliance on end-to-end differentiability and the need to relax discrete variables in order to perform inference. 

In the domain of learning-for-control, the notion of switching systems is directly related to the paradigm of model-free \gls*{HRL} \cite{barto2003recent, parr1998hierarchical}, which combines simple representations to build complex policies. Here it is useful to differentiate between two concepts of hierarchical learning, namely \emph{temporal} \cite{precup2000temporal}, and \emph{state} abstractions \cite{andre2002state}. In their seminal work \cite{sutton1999between, sutton1998intra}, the authors build on the framework of \gls*{SMDP} \cite{bradtke1995reinforcement} to learn activation/termination conditions of temporally extended actions (options) for solving discrete environments. Additionally, pioneering work in optimizing hierarchical control structures with temporally extended actions is developed in \cite{huber1997learning} and \cite{huber2000hybrid}. Recent work has focused on formulations of the \gls*{SMDP} framework that facilitate simultaneous discovery and learning of options \cite{konidaris2009skill, mankowitz2016adaptive, daniel2016probabilistic, bacon2017option, smith2018inference}.

However, the concept of state abstraction - partitioning state-action spaces into sub-regions, each governed by local dynamics and control - carries the most apparent parallels to the classical view of hybrid systems. In \cite{dietterich2000state}, a proof of convergence for RL in tabular environments with state abstraction is presented, while \cite{li2006towards} does a comprehensive study of different abstraction schemes and gives a formal definition of the problem. Furthermore, recent work has shown promising results in solving complex tasks by combining local policies, albeit while leveraging a complex neural network architecture as an upper-level policy \cite{akrour2018regularizing}.

Switching systems serve as a powerful tool in behavioral cloning. For example, \cite{calinon2010learning} combines \glspl*{HMM} with Gaussian mixture regression to represent trajectory distributions. In contrast, \cite{daniel2016probabilistic} uses a \gls*{HSMM} to learn hierarchical policies, and \cite{burke2020hybrid} introduces switching density networks for system identification and behavioral cloning. Finally, a Bayesian framework for the hierarchical policy decomposition is presented in \cite{vsovsic2017bayesian}, albeit while considering known transition dynamics.

\section{Contribution}
\label{sec:contrib}
In light of the motivation and reviewed literature from Section~\ref{sec:intro} and \ref{sec:related}, we establish here the overall contribution of our methodology and highlight the main differences that distinguish it from related approaches. 

As previously stated, this work strives to cast the problem of nonlinear optimal control into a data-driven hierarchical learning framework. Our aim is to introduce explicit structure and adopt hybrid surrogate models to avoid the opaqueness of recently popularized black-box representations. While this paradigm has been established before, our realization differs from previous attempts in two central aspects: 
\begin{itemize}[leftmargin=0.35cm]
	\item \emph{System Modeling:} This work leverages probabilistic hybrid dynamic networks as hierarchical representations of nonlinear dynamics. Contrary to a \gls*{PWARX}, \glspl*{HDBN} straightforwardly account for noise in both discrete and continuous dynamics. They also incorporate nonlinear transition boundaries, thus minimizing partitioning redundancy. Furthermore, \glspl*{HDBN} admit efficient inference methods in data-driven applications. Finally, by pursuing an abstraction over states instead of time, we circumvent the need to infer termination policies of the \gls*{SMDP} framework.
	\item \emph{Control Synthesis:} We propose a hybrid policy search approach that formulates a non-convex infinite horizon objective and optimizes a piecewise polynomial approximation of the value function with nonlinear partitioning. This approximation is used to derive stationary switching feedback controllers. In contrast, trajectory optimization and model predictive control techniques for hybrid models are often cast as sequential convex programs that assume polyhedral partitions and optimize a fixed horizon objective, yielding time-variant value functions and controls.
\end{itemize}

\section{Problem Statement}
\label{sec:problem}
Consider the discrete-time optimal control problem of a stochastic nonlinear dynamical system to be defined as an infinite horizon \gls*{MDP}. An \gls*{MDP} is defined over a state space $\mat{\mathcal{X}} \subseteq \mathds{R}^{d}$ and an action space $\mat{\mathcal{U}} \subseteq \mathds{R}^{m}$. The probability of a state transition from state $\vec{x}$ to state $\vec{x}\p$ by applying action $\vec{u}$ is governed by the Markovian time-independent density function $p(\vec{x}\p|\vec{x},\vec{u})$. The reward $r(\vec{x}, \vec{u})$ is a function of the state $\vec{x}$ and action $\vec{u}$. The state-dependent policy $\pi(\vec{u} | \vec{x})$, from which the actions are drawn, is a density determining the probability of an action $\vec{u}$ given a state $\vec{x}$. The general objective in an average-reward infinite horizon optimal control problem is to maximize the average of rewards $V^{\pi}(\vec{x}) = \lim_{T\to\infty} \frac{1}{T} \mathbb{E} \left[\sum_{t=1}^{T} r \right]$, where $V^{\pi}$ denotes as the state-value function under the policy $\pi$, starting from an initial state distribution $\mu_{1}(\vec{x})$.

\begin{figure}[t]
	\begin{minipage}{\columnwidth}
		\resizebox{\columnwidth}{!}{\input{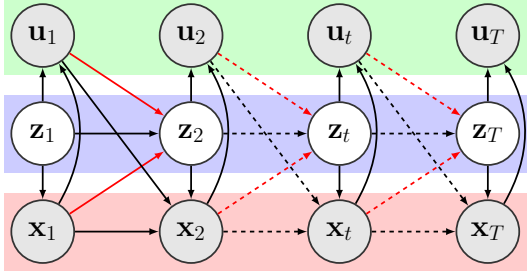}}
	\end{minipage}
	\caption{Graphical model of \acfp*{rARHMM} extended to support hybrid controls. In \acsp*{rARHMM}, the discrete state $\vec{z}$ explicitly depends on the continuous state $\vec{x}$ and action $\vec{u}$, as highlighted in red. Figure reproduced from \cite{abdulsamad2020hierarchical}.}
	\label{fig:rarhmm_pgm}
	\vspace{-0.4cm}
\end{figure}

Given the context of this work and our choice to model the system with hybrid models, we introduce to the \gls*{MDP} formulation a new hidden discrete variable $\vec{z}$, an indicator of the currently active local regime. The resulting transition dynamics can then be expressed by a factorized density function $p(\vec{x}\p, \vec{z}\p | \vec{x}, \vec{u}, \vec{z}) = p(\vec{z}\p | \vec{z}, \vec{x}, \vec{u}) p(\vec{x}\p | \vec{x}, \vec{u}, \vec{z}\p)$, which we depict as a graphical model in Figure~\ref{fig:rarhmm_pgm} and discuss in further detail in the upcoming section. In the same spirit of simplification through hierarchical modeling, we employ a mixture of switching polynomial controllers $\pi(\vec{u } | \vec{x}, \vec{z})$, associated with a piecewise polynomial value function $V^{\pi}(\vec{x}, \vec{z})$.

\section{Hybrid Dynamic Bayesian Networks}
\label{sec:hdbn}
In this section, we focus on the modeling assumptions for the stochastic switching transition dynamics $p(\vec{x}\p, \vec{z}\p | \vec{x}, \vec{u}, \vec{z})$, see Section~\ref{sec:problem}. We choose \glspl*{rARHMM} as a representation, which is a special case of \gls*{rSLDS} \cite{linderman2017recurrent}, also known as augmented \acs*{SLDS} \cite{barber2006expectation}. In contrast to \gls*{rSLDS}, an \gls*{rARHMM} lacks an observation model and directly describes the internal state up to an additive noise process. We extend \glspl*{rARHMM} to support exogenous and endogenous inputs in order to simulate the open- and closed-loop behaviors of driven dynamics. Figure~\ref{fig:rarhmm_pgm} depicts the corresponding graphical model, which closely resembles the graph of a \gls*{PWARX}.

An \gls*{rARHMM} with $K$ regions models the trajectory of a dynamical system as follows. The initial continuous state $\vec{x}_{1} \in \mathds{R}^{d}$ and continuous action $\vec{u}_{1} \in \mathds{R}^{m}$ are drawn from a pair of Gaussian and conditional Gaussian distributions\footnote{We parameterize all Gaussian distributions by their precision matrices instead of the more common definition with covariances.}, respectively. The initial discrete state $\vec{z}_{1}$ is a random vector modeled by a categorical density parameterized by $\vec{\varphi}$
\begin{gather*}
	\vec{z}_{1} \sim \Cat(\vec{\varphi}), ~ \vec{x}_{1} \sim \N(\vec{\mu}_{\vec{z}_{1}}, \mat{\Omega}_{\vec{z}_{1}}), \\
	\vec{u}_{1} \sim \N(\mat{K}_{\vec{z}_{1}} \phi(\vec{x}_{1}), \mat{\Delta}_{\vec{z}_{1}}).
\end{gather*}
The transition of the continuous state $\vec{x}_{t+1}$ and actions $\vec{u}_{t}$ are modeled by affine-Gaussian dynamics
\begin{alignat*}{2}
	\vec{x}_{t+1} & \! = \! \mat{A}_{\vec{z}_{t+1}} \vec{x}_{t} \! + \mat{B}_{\vec{z}_{t+1}} \vec{u}_{t} \! + \vec{c}_{\vec{z}_{t+1}} \! + \vec{\lambda}_{t}, ~ &  & \vec{\lambda}_{t} \sim \N(\vec{0}, \mat{\Lambda}_{\vec{z}_{t+1}}), \\ 
	\vec{u}_{t}   & \! = \! \mat{K}_{\vec{z}_{t}} \phi(\vec{x}_{t}) + \vec{\delta}_{t}, ~                                                           &  & \vec{\delta}_{t} \sim \N(\vec{0}, \mat{\Delta}_{\vec{z}_t}),                  
\end{alignat*}
where $(\mat{A}, \mat{B}, \vec{c}, \mat{K}, \mat{\Omega}, \mat{\Lambda}, \mat{\Delta} )$ are matrices and vectors of appropriate dimensions with respect to $\vec{x}$ and $\vec{u}$. $\phi(\vec{x})$ are polynomial state features of arbitrary degree. 

The discrete transition probability $p(\vec{z}_{t+1} | \vec{z}_{t}, \vec{x}_{t}, \vec{u}_{t})$ is governed by $K$ categorical distributions parameterized by a state-action dependent multi-class logit link function $f$ \cite{murphy2012machine}
\begin{align}
	\label{eq:trans_probs}
	\chi_{ij} \! = \! p(\vec{z}_{t+1}\!=\!j | \vec{z}_{t}\!=\!i, \vec{x}_{t}, \vec{u}_{t}) \footnotemark \! \propto \! \exp \left(f(\vec{x}_{t}, \vec{u}_{t}; \vec{\omega}_{ij}) \right),
\end{align}
\footnotetext{We abuse notation slightly by sometimes using $\vec{z}$ to refer to the discrete state index instead of treating it as a one-hot vector.}where $f$ may have any type of features in ($\vec{x}, \vec{u})$. The vectors $\vec{\omega}_{ij}$ parameterize the discrete transition probabilities for all transition combinations $i \rightarrow j~ \forall {i,j} \in [1, \, K]$. Figure~\ref{fig:rarhmm_example} depicts realizations of different logit link functions leading to various state space partitioning.

The remainder of this paper focuses on using these hybrid models in three scenarios:
\begin{itemize}
	\item An \emph{open-loop} setting that treats the control $\vec{u}$ as an exogenous input is used for automatically identifying nonlinear systems via decomposition into continuous and discrete switching dynamics.
	\item A \emph{closed-loop} setting that assumes the control $\vec{u}$ to originate from a nonlinear controller. We show that this setting can simultaneously decompose dynamics and control in a behavioral cloning scenario.
	\item A \emph{reinforcement learning} setting where we develop a model-based hybrid policy search algorithm to learn switching controllers for general nonlinear systems.
\end{itemize}

\section{Bayesian Inference of Hybrid Models}
\label{sec:sysid}
In this section, we sketch the outline of an expectation-maximization/Baum-Welch algorithm \cite{baum1970maximization, dempster1977maximum, rabiner1989tutorial} for inferring the parameters of an \gls*{rARHMM} given time-series observations. The resulting algorithm can be used two-fold. First, it can be applied to automatically identify hybrid models and approximate the open-loop dynamics of nonlinear systems given state-action observations. Second, it can clone the closed-loop behavior of a nonlinear controller and decompose it into a set of local experts.

Our developed approach is related in some aspects to the Baum-Welch algorithms proposed in \cite{bengio1995input} and \cite{daniel2016probabilistic}. However, we introduce suitable priors over all parameters and derive a \gls*{MAP} technique with a stochastic maximization step and hyperparameter optimization. In our experience, the priors significantly regularize the sensitivity of \gls*{EM} with respect to the initial point, making it less prone to getting stuck in bad local minima.

Moreover, a good prior specification is crucial in small data regimes since a vague prior may dominate the predictive posterior and effectively cause under-fitting. We implement a hyperparameter optimization scheme that elevates this concern by optimizing the prior parameters via empirical Bayes \cite{maritz1989empirical}, thus attenuating the prior influence and improving the predictive performance significantly. 

\subsection{Maximum A Posteriori Optimization}
\label{subsec:em}
Consider again the \gls*{rARHMM} in Figure~\ref{fig:rarhmm_pgm} where the continuous state $\vec{x}$ and action $\vec{u}$ are observed variables, while the $K$-region indicators $\vec{z}$ are hidden. To infer the model parameters, we assume a dataset of $N$ state-action trajectories $\mat{\mathcal{D}} = \{ \mat{\mathcal{D}}^{n} \}^{N}_{n=1} = \{ \mat{X}^{n}, \mat{U}^{n} \}^{N}_{n=1}$, each of length $T$, where $(\mat{X}^{n}$, $\mat{U}^{n}, \mat{Z}^{n})$ represent the time concatenation of an entire trajectory $(\vec{x}^{n}_{1:T}$, $\vec{u}^{n}_{1:T}, \vec{z}^{n}_{1:T})$.

The objective corresponding to system identification and behavioral cloning can be cast as a maximization problem of the log-posterior probability of the observations $\{ (\mat{X}^{n}, \mat{U}^{n}) \}^{N}_{n=1}$, with respect to the free parameter set $\vec{\theta} = \{\vec{\varphi}, \vec{\mu}_{k}, \mat{\Omega}_{k}, \mat{A}_{k}, \mat{B}_{k}, \vec{c}_{k}, \mat{\Lambda}_{k}, \mat{K}_{k}, \mat{\Delta}_{k}, \vec{\omega}_{ik} \}_{i,k=1}^{K}$
\begin{equation} \label{eq:log_prob}
	\vec{\theta}_{\text{MAP}} := \argmax_{\vec{\theta}} ~ \log \prod_{n=1}^{N} \sum_{\vec{z}^{n}} p(\mat{\mathcal{D}}^{n}, \mat{Z}^{n} | \vec{\theta}) p(\vec{\theta} | \vec{h}),
\end{equation}
where $p(\mat{\mathcal{D}}^{n}, \mat{Z}^{n} | \vec{\theta})$ is the complete-data likelihood of a single trajectory and factorizes according to
\begin{equation} \label{eq:traj_lklhd}
	p(. | \vec{\theta}) =
	\begin{aligned}[t]
		& p(\vec{z}_{1}^{n} | \vec{\varphi}) p(\vec{x}_{1}^{n} | \vec{\mu}_{\vec{z}_{1}^{n}}, \mat{\Omega}_{\vec{z}_{1}^{n}}) p(\vec{u}^{n}_{1} | \vec{x}_{1}^{n}, \mat{K}_{\vec{z}_{1}^{n}}, \mat{\Delta}_{\vec{z}_{1}^{n}}) \\
		& \times \prod_{t=2}^{T} p(\vec{x}_{t}^{n} | \vec{x}_{t-1}^{n}, \vec{u}_{t-1}^{n}, \mat{A}_{\vec{z}_{t}^{n}}, \mat{B}_{\vec{z}_{t}^{n}}, \vec{c}_{\vec{z}_{t}^{n}}, \mat{\Lambda}_{\vec{z}_{t}^{n}}) \\ 
		& \times \prod_{t=2}^{T} p(\vec{z}_{t}^{n} | \vec{z}_{t-1}^{n}, \vec{x}_{t-1}^{n}, \vec{u}_{t-1}^{n}, \vec{\omega}) \\ 
		& \times \prod_{t=2}^{T} p(\vec{u}_{t}^{n} | \vec{x}_{t}^{n}, \mat{K}_{\vec{z}_{t}^{n}}, \mat{\Delta}_{\vec{z}_{t}^{n}}),
	\end{aligned}
\end{equation}
and $p(\vec{\theta} | \vec{h})$ is the factorized parameter prior
\begin{equation*} 
	p(\vec{\theta} | \vec{h}) = 
	\begin{aligned}[t]
		& p(\vec{\varphi}) \prod_{k=1}^{K} p(\vec{\mu}_{k} | \mat{\Omega}_{k}) p(\mat{\Omega}_{k}) \\ 
		& \times \prod_{k=1}^{K}  p(\mat{A}_{k} | \mat{\Lambda}_{k}) p(\mat{B}_{k} | \mat{\Lambda}_{k}) p(\vec{c}_{k} | \mat{\Lambda}_{k}) p(\mat{\Lambda}_{k}) \\
		& \times \prod_{k=1}^{K} p(\mat{K}_{k} | \mat{\Delta}_{k}) p(\mat{\Delta}_{k}) \prod_{i=1}^{K} \prod_{k=1}^{K} p(\vec{\omega}_{ik}).
	\end{aligned}
\end{equation*}

We choose all priors to be conjugate or semi-conjugate with respect to their likelihoods. Therefore, we place a \gls*{NW} prior on the initial state distribution $(\vec{\mu}_{k}, \mat{\Omega}_{k}) \sim \NW(\vec{0}, \kappa_{0}, \mat{\Psi}_{0}, \nu_{0})$, and a \gls*{MNW} on the affine transition dynamics $(\mat{A}_{k}, \mat{B}_{k}, \mat{c}_{k}, \mat{\Lambda}_{k}) \sim \MNW(\mat{0}, \mat{R}_{0}, \mat{\Phi}_{0}, \rho_{0})$. The initial discrete state takes a Dirichlet prior $\vec{\varphi} \sim \Dir(\vec{\tau}_{0})$, while the logit link function parameters are governed by a non-conjugate zero-mean Gaussian prior with diagonal precision $\vec{\omega}_{ik} \sim \N(\vec{0}, \alpha \mat{I})$. Finally, we place a separate matrix-normal-Wishart prior on the conditional action likelihood $(\mat{K}_{k}, \mat{\Delta}_{k}) \sim \MNW(\mat{0}, \mat{S}_{0}, \mat{\Gamma}_{0}, \varepsilon_{0})$. The quantities $(\kappa_{0}, \mat{\Psi}_{0}, \nu_{0}, \mat{R}_{0}, \mat{\Phi}_{0}, \rho_{0}, \vec{\tau}_{0}, \alpha, \mat{S}_{0}, \mat{\Gamma}_{0}, \varepsilon_{0})$ are hyperparameters aggregated into the hyperparameter set $\vec{h}$. 

The choice of priors is not restricted to these distributions. Depending on modeling assumptions, one can assume dynamics with diagonal noise matrices and pair them with gamma distribution priors. Moreover, if the system is known to have a state-independent noise process, the $K$ Wishart and gamma priors can be \emph{tied} across components, leading to a more structured representation.

\subsection{Baum-Welch Expectation-Maximization}
On closer examination of Equations~\eqref{eq:log_prob} and \eqref{eq:traj_lklhd}, we observe that the optimization problem is non-convex with multiple local optima since the complete-data likelihood $\prod_{n=1}^{N} p(\mat{\mathcal{D}}^{n}, \mat{Z}^{n} | \vec{\theta})$ can follow complex multi-modal densities. Another technical difficulty is the summation over all possible trajectories of the hidden variables $\mat{Z}^{n}$, which is of computational complexity $\mathcal{O}(NK^{T})$ and is intractable in most cases. Expectation-maximization algorithms overcome the latter problem by introducing a variational posterior distribution over the hidden variables $q(\mat{Z}^{n})$ and deriving a lower bound on the complete log-probability function 
\begin{multline} \label{eq:em_bound}
	\log \prod_{n=1}^{N} \sum_{\vec{z}^{n}} p(\mat{\mathcal{D}}^{n}, \mat{Z}^{n}, \vec{\theta} | \vec{h}) \\
	\geq \sum_{n=1}^{N} \sum_{\vec{z}^{n}} q(\mat{Z}^{n}) \log \frac{p(\mat{\mathcal{D}}^{n}, \mat{Z}^{n}, \vec{\theta} | \vec{h})}{q(\mat{Z}^{n})}.
\end{multline}

We find a point estimate $\vec{\theta}_{\text{MAP}}$ by following a modified scheme of EM, alternating between an \gls*{E-step}, in which the lower bound in Equation~\eqref{eq:em_bound} is maximized with respect to the variational distributions $q(\mat{Z}^{n})$ given a parameter estimate $\hvec{\theta}$, a \gls*{M-step}, that updates $\vec{\theta}$ given $({\hat q}(\mat{Z}^{n}), \hvec{h})$, and finally, an \gls*{EB-step} that updates $\vec{h}$ given $({\hat q}(\mat{Z}^{n}), \hvec{\theta})$. A sketch of the overall iterative procedure is presented in Algorithm~\ref{alg:em_algo}.

\subsubsection{Exact Expectation Step}
Maximizing the lower bound with respect to $q(\mat{Z}^{n})$ is determined by reformulating Equation~\eqref{eq:em_bound}
\begin{align*}
	L & \! = \! \sum_{n=1}^{N} \sum_{\vec{z}^{n}} q(\mat{Z}^{n}) \log \frac{p(\mat{\mathcal{D}}^{n}, \mat{Z}^{n}, \vec{\theta} | \vec{h})}{q(\mat{Z}^{n})} \\
	& \! = \! \sum_{n=1}^{N} \log p(\mat{\mathcal{D}}^{n} \!, \vec{\theta} | \vec{h})
	+ \sum_{n=1}^{N} \sum_{\vec{z}^{n}} q(\mat{Z}^{n}) \log \frac{p(\mat{Z}_{n}| \mat{\mathcal{D}}^{n}, \vec{\theta})}{q(\mat{Z}^{n})} \\
	& \! = \! \sum_{n=1}^{N} \! \log p(\mat{\mathcal{D}}^{n} \!, \vec{\theta} | \vec{h}) \! - \! \sum_{n=1}^{N} \kl (q(\mat{Z}^{n}) \, || \, p(\mat{Z}^{n} | \mat{\mathcal{D}}^{n} \!, \vec{\theta})).
\end{align*}

This form of the lower bound implies that the optimal variational distribution ${\hat q}(\mat{Z}^{n})$ minimizes the \gls*{KL} \cite{kullback1951information}, meaning
\begin{equation} \label{eq:e_step}
	{\hat q}(\mat{Z}^{n}) =  p(\mat{Z}^{n} | \mat{\mathcal{D}}^{n}, \vec{\theta}) = p(\vec{z}^{n}_{1:T} | \vec{x}_{1:T}^{n}, \vec{u}_{1:T}^{n}, \vec{\theta}).
\end{equation}

This update tightens the bound if the posterior model ${\hat q}(\mat{Z}^{n})$ belongs to the same family of the true posterior \cite{beal2003variational}. Notice that the \gls*{E-step} is independent of the prior $p(\vec{\theta})$. Moreover, Equation~\eqref{eq:e_step} indicates that the \gls*{E-step} reduces to the computation of the smoothed marginals $p(\vec{z}_{t}^{n} | \vec{x}_{1:T}^{n}, \vec{u}_{1:T}^{n}, \hvec{\theta})$ under the current parameter estimate $\hvec{\theta}$. Following \cite{baum1970maximization} and \cite{murphy2012machine}, we derive a two-filter algorithm, which enables closed-form and exact inference by splitting the smoothed marginals into a forward and backward message\footnote{We briefly drop the dependency on $\hvec{\theta}$ for an uncluttered notation while deriving the forward-backward recursions.}
\begin{align*}
	\vec{\gamma}_{t}^{n}(k) & \! = p(\vec{z}_{t}^{n} \! = \! k | \vec{x}_{1:T}^{n}, \vec{u}_{1:T}^{n}) \\
	& \! \propto p(\vec{z}_{t}^{n} \! = \! k | \vec{x}_{1:t}^{n}, \vec{u}_{1:t}^{n}) p(\vec{x}_{t+1:T}^{n}, \vec{u}_{t+1:T}^{n} | \vec{z}_{t}^{n} \! = \! k, \vec{x}_{t}^{n}, \vec{u}_{t}^{n}) \\
	& \! = \alpha_{t}^{n}(k) \beta_{t}^{n}(k),
\end{align*}
where $\alpha_{t}^{n}(k) = p(\vec{z}_{t}^{n} \! = \! k | \vec{x}_{1:t}^{n}, \vec{u}_{1:t}^{n})$ is the message which computes the filtered marginals via a forward recursion 
\begin{equation*}
	\alpha_{t}^{n}(k) \propto 
	\begin{aligned}[t]
		& p(\vec{x}_{t}^{n} | \vec{x}_{t-1}^{n}, \vec{u}_{t-1}^{n}, \vec{z}_{t}^{n} \!=\! k) p(\vec{u}_{t}^{n} | \vec{x}_{t}^{n}, \vec{z}_{t}^{n} \! = \! k) \\ 
		& \times \sum^{K}_{j=1} p(\vec{z}_{t}^{n} = k | \vec{z}_{t-1}^{n} \! = \! j, \vec{x}_{t-1}^{n}, \vec{u}_{t-1}^{n}) {\alpha}_{t-1}^{n}(j),
	\end{aligned}
\end{equation*}
and $\beta_{t}^{n}(k) = p(\vec{x}_{t+1:T}^{n} | \vec{z}_{t}^{n}=k, \vec{x}_{t}^{n}, \vec{u}_{t}^{n})$ is the backward message that performs smoothing by computing the conditional likelihood of future evidence
\begin{equation*}
	\beta_{t}^{n}(k) \! =
	\begin{aligned}[t]
		\! \sum_{j=1}^{K} & \beta_{t+1}^{n}(j) p(\vec{z}_{t+1}^{n} \! = \! j | \vec{z}_{t}^{n} \! = \! k, \vec{x}_{t}^{n}, \vec{u}_{t}^{n}) \\
		& \hspace{-0.35cm} \times \! p(\vec{x}_{t+1}^{n} | \vec{x}_{t}^{n}, \vec{u}_{t}^{n}, \vec{z}_{t+1}^{n} \! = \! j) p(\vec{u}_{t+1}^{n} | \vec{x}_{t+1}^{n}, \vec{z}_{t+1}^{n} \! = \! j).
	\end{aligned}
\end{equation*}

Additionally, by combining both forward and backward messages, we can compute the two-slice smoothed marginals $p(\vec{z}_{t}^{n}, \vec{z}_{t+1}^{n} | \vec{x}_{1:T}^{n}, \vec{u}_{1:T}^{n})$ which will be useful during the maximization and empirical Bayes steps
\begin{align*}
	\xi_{t,t+1}^{n}(i,j) & = p(\vec{z}_{t}^{n} \! = \! i, \vec{z}_{t+1}^{n} \! = \! j | \vec{x}_{1:T}^{n}, \vec{u}_{1:T}^{n}) \\ 
	& \propto
	\begin{aligned}[t]
		& p(\vec{x}_{t+1}^{n} | \vec{x}_{t}^{n}, \vec{u}_{t}^{n}, \vec{z}_{t+1}^{n} \! = \! j) p(\vec{u}_{t+1}^{n} | \vec{x}_{t+1}^{n}, \vec{z}_{t+1}^{n} \! = \! j) \\
		& \times {\alpha}_{t}^{n}(i) p(\vec{z}_{t+1}^{n} \! = \! j | \vec{z}_{t}^{n} \! = \! i, \vec{x}_{t}^{n}, \vec{u}_{t}^{n}) {\beta}_{t+1}^{n}(j).
	\end{aligned}
\end{align*}

\subsubsection{Stochastic Maximization Step}
After performing the \gls*{E-step} and computing the smoothed posteriors, we are able to evaluate the lower bound and maximize it with respect to $\vec{\theta}$ given $({\hat q}(\mat{Z}^{n}), \hvec{h})$. 

By plugging Equations~\eqref{eq:traj_lklhd} and \eqref{eq:e_step} into \eqref{eq:em_bound}, leveraging conditional independence, and disregarding terms independent of $\vec{\theta}$, we arrive at the expected complete log-probability function $Q(\vec{\theta}, \vec{\gamma}, \vec{\xi}, \hvec{h})$
\begin{align} \label{eq:m_step}
	Q & = \sum_{n=1}^{N} \sum_{\vec{z}^{n}} \hat{q}(\mat{Z}^{n}) \log p(\mat{\mathcal{D}}^{n}, \mat{Z}^{n}, \vec{\theta} | \hvec{h}) \nonumber \\
	& = 
	\begin{aligned}[t]
		& \log p(\vec{\theta} | \hvec{h}) \! + \! \sum_{k=1}^{K} \! \sum_{n=1}^{N} \gamma^{n}_{1} \Big[\log \varphi_{k} + \log \N(\vec{x}_{1}^{n} | \vec{\mu}_{k}, \mat{\Omega}_{k}) \Big] \nonumber \\
		& + \sum_{k=1}^{K} \! \sum_{n=1}^{N} \! \sum_{t=2}^{T} \gamma^{n}_{t} \log \N(\vec{x}_{t}^{n} | \mat{A}_{k} \vec{x}_{t-1}^{n} \! + \mat{B}_{k} \vec{u}_{t-1}^{n} \! + \vec{c}_{k}, \mat{\Lambda}_{k}) \nonumber \\
		& + \sum_{k=1}^{K} \! \sum_{n=1}^{N} \! \sum_{t=1}^{T} \gamma^{n}_{t} \log \N(\vec{u}_{t}^{n} | \mat{K}_{k} \phi(\vec{x}_{t-1}^{n}), \mat{\Delta}_{k}) \nonumber \\
		& + \sum_{i=1}^{K} \! \sum_{j=1}^{K} \! \sum_{n=1}^{N} \! \sum_{t=2}^{T} \xi^{n}_{t-1, t} \log \chi_{ij} (\vec{x}^{n}_{t-1}, \vec{u}^{n}_{t-1}, \vec{\omega}_{ij}).
	\end{aligned}
\end{align}
The function $Q$ is non-convex in $\vec{\omega}$ when a nonlinear logit link function $f(., \vec{\omega})$ is chosen as an embedding for the transition probability $\chi$, see Equation~\eqref{eq:trans_probs}. In that case, stochastic optimization is recommended \cite{robbins1951stochastic} as batched noisy gradient estimates allow the algorithm to escape shallow local minima and reduce the computational cost that comes with evaluating the gradients for all data instances. 

Consequently, when implementing the \gls*{M-step}, we apply stochastic optimization on the transition parameters $\vec{\omega}$. We use a stochastic gradient ascent direction with an adaptive learning rate $\varepsilon$  and batch size $M$ \cite{robbins1951stochastic}
\begin{gather*}
	\vec{\omega}^{(l + 1)} = \vec{\omega}^{l} + \frac{\varepsilon}{M} \sum_{m=1}^{M} \nabla_{\vec{\omega}} Q^{(m)} \left. \right\vert_{\vec{\omega} = \vec{\omega}^{l}}, \\
	\nabla_{\vec{\omega}} Q^{(m)} = 
	\begin{aligned}[t]
		\nabla_{\vec{\omega}} \Big[ & \log p(\vec{\omega} | \alpha) \\
		& + \sum_{i=1}^{K} \sum_{j=1}^{K} \xi^{(m)} \log \chi_{ij}(\vec{x}^{(m)}, \vec{u}^{(m)}, \vec{\omega}_{ij}) \Big].
	\end{aligned}
\end{gather*}

\removealgorithmerror
\SetEndCharOfAlgoLine{}
\RestyleAlgo{boxruled}
\begin{algorithm}[!t]
	\linespread{1.25}\selectfont
	\SetKwInput{Input}{input}
	\SetKwInput{Output}{output}
	\SetKwInput{Initialize}{initialize}
	\SetKwRepeat{Do}{do}{while}
	
	\Input{$\mat{\mathcal{D}} = \{\mat{X}^{n}, \mat{U}^{n} \}_{n=1}^{N}, \vec{h}, K$}
	\Initialize{$\hvec{\theta} \sim p(\vec{\theta} | \vec{h}), \hvec{h} \leftarrow \vec{h}$}
	
	\While{$\log p(\mat{\mathcal{D}}, \vec{\theta} | \vec{h})$ \normalfont not converged}
	{
		\tcp{Expectation step}
		\For{$n \gets1$ \KwTo $N$}
		{
			$\vec{\alpha}^{n}, \vec{\beta}^{n} \leftarrow $ \bf{ForwardBackward}$(\mat{X}^{n}, \mat{U}^{n}, \hvec{\theta})$ \\
			$\vec{\gamma}^{n}, \vec{\xi}^{n} \leftarrow $ \bf{SmoothedPosteriors}$(\vec{\alpha}^{n}, \vec{\beta}^{n}, \hvec{\theta})$
		}
		\tcp{Maximization step}
		$\hvec{\theta}$ $\leftarrow$ \bf{Maximize} $Q(\vec{\theta}, \vec{\gamma}, \vec{\xi}, \hvec{h})$ \\
		\tcp{Empirical Bayes}
		$\hvec{h} \leftarrow \hvec{h} + \varrho \, \nabla_{\vec{h}} Q \left. \right\vert_{\vec{h}=\hvec{h}}$
	}
	\Output{$\hvec{\theta}$}
	
	\linespread{1.0}\selectfont
	\caption{Expectation-Maximization for System Identification and Behavioral Cloning}
	\label{alg:em_algo}
\end{algorithm}

For the parameters with conjugate priors, we derive closed-form optimality conditions. Effectively, we derive the posterior distribution via Bayes' rule and take the mode of each posterior density for a \gls*{MAP} estimate update. 

By considering only relevant terms, we write the \gls*{MAP} of the initial gating parameter $\vec{\varphi}$ as
\begin{equation*}
	\max_{\vec{\varphi}} \quad \log \, \Dir(\vec{\varphi} | \hvec{\tau}_{0}) + \sum_{k=1}^{K} \! \sum_{n=1}^{N} \gamma^{n}_{1} \log \varphi_{k},
\end{equation*}
while the estimate of the initial state parameters $(\vec{\mu}_{k}, \mat{\Omega}_{k})$ can be decoupled for each $k$ as follows
\begin{align*}
	\max_{(\vec{\mu}, \mat{\Omega})_{k}} \quad & \log \, \NW(\vec{\mu}_{k}, \mat{\Omega}_{k} | (\mat{0}, \hat{\kappa}_{0}, \hmat{\Psi}_{0}, \hat{\nu}_{0})_{k}) \\ 
	& + \sum_{n=1}^{N} \gamma^{n}_{1} \log \N(\vec{x}_{1}^{n} | \vec{\mu}_{k}, \mat{\Omega}_{k}).
\end{align*}

Analogously, the \gls*{MAP} of the dynamics parameter $(\mat{A}_{k}, \mat{B}_{k}, \mat{c}_{k}, \mat{\Lambda}_{k})$ is also decoupled to $k$ optimizations
\begin{align*}
	\max_{(\mat{A}, \mat{B}, \vec{c}, \mat{\Lambda})_{k}} \quad & \log \, \MNW(\mat{A}_{k}, \mat{B}_{k}, \vec{c}_{k}, \mat{\Lambda}_{k} | (\mat{0}, \hmat{R}_{0}, \hmat{\Phi}_{0}, \hat{\nu}_{0})_{k}) \\
	& \hspace{-0.70cm} + \sum_{n=1}^{N} \! \sum_{t=2}^{T} \gamma^{n}_{t} \log \N(\vec{x}_{t}^{n} | \mat{A}_{k} \vec{x}_{t-1}^{n} \! + \mat{B}_{k} \vec{u}_{t-1}^{n} \! + \vec{c}_{k}, \vec{\Lambda}_{k}),
\end{align*}
and, finally, to learn closed-loop behavior, we can infer the controller parameters $(\mat{K}_{k}, \mat{\Delta}_{k})$
\begin{align*}
	\max_{(\mat{K}, \mat{\Delta})_{k}} \quad & \log \, \MNW(\mat{K}_{k}, \mat{\Delta}_{k} | (\mat{0}, \hmat{S}_{0}, \hmat{\Gamma}_{0}, \hat{\varepsilon}_{0})_{k}) \\
	& \! + \! \sum_{n=1}^{N} \! \sum_{t=1}^{T} \gamma^{n}_{t} \log \N(\vec{u}_{t}^{n} | \mat{K}_{k} \phi(\vec{x}^{n}_{t}), \vec{\Delta}_{k}).
\end{align*}
Due to space constraints, we will refrain from stating the explicit solution for these optimization problems. Instead, we provide the general outline of how to compute these posteriors and their modes based on the unified notation for exponential family distributions in Appendix~\ref{app:expo} and \ref{app:posteriors}.

\subsubsection{Approximate Empirical Bayes}
Inference techniques that leverage data-independent assumptions run the risk of prior miss-specification. In our MAP approach, the priors are weakly informative and carry little information. Their main purpose is to regularize greedy updates that might lead to premature convergence. However, when there is little data, the priors, especially those on the precision matrices, %
may dominate the posterior probability, leading to over-regularization and under-fitting of the objective. Empirical Bayes approaches remedy this issue by integrating out the parameters $\vec{\theta}$ and optimizing the marginal likelihood with respect to the hyperparameters $\vec{h}$ \cite{maritz1989empirical}. In our setting, marginalizing all hidden quantities does not admit a closed-form formula. An approximate approach to empirical Bayes is to interleave the E- and M-steps with hyperparameter updates that optimize the lower bound given an estimate of parameters $\hvec{\theta}$ and a step size $\varrho$
\begin{equation*}
	\vec{h}^{(l + 1)} = \vec{h}^{(l)} + \varrho \, \nabla_{\vec{h}} Q \left. \right\vert_{\vec{h}=\vec{h}^{l}},
\end{equation*}
where the gradient of $Q$ with respect to $\vec{h}$ reduces to
\begin{equation*}
	\nabla_{\vec{h}} Q = \nabla_{\vec{h}} \log p(\hvec{\theta} | \vec{h}).
\end{equation*}

\section{Reinforcement Learning via Hybrid Models}
\label{sec:rl}
The last sections focused on the system modeling aspect and how to use hybrid surrogate models to approximate nonlinear dynamics. Now we turn our attention to the problem of using these models to synthesize structured controllers for general nonlinear dynamical systems. One possible approach is to use the learned hybrid models and apply the classical hybrid control methods, which we have reviewed Section~\ref{sec:related}. However, as discussed earlier, these methods suffer from several drawbacks. On the one hand, they rely a polyhedral partitioning of the space. This limitation is severe because it often leads to an explosion in the number of partitions. On the other hand, these methods are often focused on computationally expensive trajectory-centric model predictive control. This class of controllers is disadvantageous in applications that require fast reactive feedback signals with broad coverage over the state-action space.

In this section, we address these points and present an infinite horizon stochastic optimization technique that incorporates the structure of hybrid models. This approach can deal with \glspl*{rARHMM} with arbitrary non-polyhedral partitioning and synthesizes stationary piecewise polynomial controllers. Our algorithm extends the step-based formulation of \gls*{REPS} \cite{peters2010relative, van2015learning, belousov2017f} by explicitly accounting for the discrete-continuous mixture state variables $(\vec{x}, \vec{z})$. Our approach, hybrid REPS (Hb-REPS), leverages the state-action-dependent nonlinear switches $p(\vec{z}\p | \vec{z}, \vec{x}, \vec{u})$ as a task-independent upper-level coordinator to a mixture of $K$ lower-level policies $\pi(\vec{u} | \vec{x}, \vec{z})$. While the proposed approach shares many features with \cite{daniel2016probabilistic}, our formulation relies on a state-abstraction representation of hybrid models and embeds the hierarchical model structure into the optimization problem in order to learn a hierarchy over the global value function. In contrast, \cite{daniel2016probabilistic} operates in the framework of semi-Markov decision processes and optimizes a mixture over termination and feedback policies without considering the existence of a hierarchical structure in the space of dynamics and value functions. For more details on differences between state- and time-abstractions, refer to Section~\ref{sec:related}.

\subsection{Infinite-Horizon Stochastic Optimal Control}
\label{sec:hbreps}
In the \gls*{REPS} framework, an optimal control problem is presented as an iterative trust-region optimization for a discounted average-reward objective under a stationary state-action distribution $\pi(\vec{u} | \vec{x}, \vec{z}) \mu(\vec{x}, \vec{z})$, Equation~\eqref{eq:reps_obj}. The trust-region is formulated as a \gls*{KL} \cite{kullback1951information}, Equation~\eqref{eq:reps_kl}. Its purpose is to regularize the search direction and limit information loss between iterations. The REPS formulation explicitly incorporates a dynamics consistency constraint, Equation~\eqref{eq:reps_dynamics}, that describes the evolution of the stochastic state of the system. The following optimization problem is solved during a single iteration of hybrid REPS
\begin{subequations}
	\begin{alignat}{3}
		& \max_{\pi, \mu} & \quad & J = \sum_{\vec{z}} \iint r(\vec{x}, \vec{u}) \pi(\vec{u} | \vec{x}, \vec{z}) \mu(\vec{x}, \vec{z}) \dif \vec{x} \dif \vec{u}, \label{eq:reps_obj} \\
		& \st             & \quad & \mu(\vec{x}\p, \vec{z}\p) = (1 - \vartheta) \mu_{1}(\vec{x}\p, \vec{z}\p) \label{eq:reps_dynamics} \\
		&                 & \quad & + \vartheta \sum_{\vec{z}} \iint \pi(\vec{u} | \vec{x}, \vec{z}) \mu(\vec{x}, \vec{z}) p(\vec{x}\p, \vec{z}\p | \vec{x}, \vec{u}, \vec{z}) \dif \vec{u} \dif \vec{x}, \nonumber \\
		& \quad           & \quad & \kl (\pi(\vec{u} | \vec{x}, \vec{z}) \mu(\vec{x}, \vec{z}) \, || \, q(\vec{x}, \vec{u}, \vec{z})) \leq \epsilon, \label{eq:reps_kl} \\
		& \quad           & \quad & \sum_{\vec{z}} \iint \pi(\vec{u} | \vec{x}, \vec{z}) \mu(\vec{x}, \vec{z}) \dif \vec{x} \dif \vec{u} = 1, \label{eq:reps_norm}
	\end{alignat}
\end{subequations}
where $\mu(\vec{x}, \vec{z})$ is the stationary mixture distribution, $q(\vec{x}, \vec{u}, \vec{z})$ is the trust-region reference distribution, and the constraint in Equation~\eqref{eq:reps_norm} guarantees the normalization of the state-action distribution. The factor $1 - \vartheta$, $\vartheta \in [0, 1)$, represents the probability of an infinite process to reset to an initial distribution $\mu_{1}(\vec{x}, \vec{z})$. The notion of resetting is necessary to ensure ergodicity of the closed-loop Markov process and allows the interpretation of $\vartheta$ as a discount factor and regularization of the \gls*{MDP} \cite{puterman2014markov, belousov2017f}. 

\subsection{Optimality Conditions and Dual Optimization}
To solve the trust-region optimization in Equations~\eqref{eq:reps_obj}-\eqref{eq:reps_norm}, we start by constructing the Lagrangian of the primal \cite{boyd2004convex}
\begin{equation*}
	\mathcal{L} =
	\begin{aligned}[t]
		& \sum_{\vec{z}} \iint r(\vec{x}, \vec{u}) p(\vec{x}, \vec{u}, \vec{z}) \dif \vec{x} \dif \vec{u} \\
		& + \sum_{\vec{z}\p} \int V(\vec{x}\p, \vec{z}\p) \Big[ - \int p(\vec{x}\p, \vec{z}\p, \vec{u}\p) \dif \vec{u}\p \\
		& + (1 - \vartheta) \sum_{\vec{z}} \iint p(\vec{x}, \vec{u}, \vec{z}) \mu_{1}(\vec{x}\p, \vec{z}\p) \dif \vec{x} \dif \vec{u} \\ 
		& + \vartheta \sum_{\vec{z}} \iint p(\vec{x}, \vec{u}, \vec{z}) p(\vec{x}\p, \vec{z}\p | \vec{x}, \vec{u}, \vec{z}) \dif \vec{x} \dif \vec{u} \Big] \dif \vec{x}\p \\
		& + \lambda \Big[ 1 - \sum_{\vec{z}} \iint p(\vec{x}, \vec{u}, \vec{z}) \dif \vec{u} \dif \vec{x} \Big] \\
		& + \eta \Big[ \epsilon - \sum_{\vec{z}} \iint p(\vec{x}, \vec{u}, \vec{z}) \log \frac{p(\vec{x}, \vec{u}, \vec{z})}{q(\vec{x}, \vec{u}, \vec{z})} \dif \vec{x} \dif \vec{u} \Big],
	\end{aligned}
\end{equation*}
where we use $p(\vec{x}, \vec{u}, \vec{z}) = \mu(\vec{x}, \vec{z}) \pi(\vec{u} | \vec{x}, \vec{z})$ for convenience and leverage the following identities
\begin{align*}
	\mu(\vec{x}, \vec{z}) & = \int p(\vec{x}, \vec{u}, \vec{z}) \dif \vec{u}, \\
   	\mu_{1}(\vec{x}\p, \vec{z}\p) & = \sum_{\vec{z}} \iint p(\vec{x}, \vec{u}, \vec{z}) p_{1}(\vec{x}\p, \vec{z}\p | \vec{x}, \vec{u}, \vec{z}) \dif \vec{x} \dif \vec{u} \\
	& = \sum_{\vec{z}} \iint p(\vec{x}, \vec{u}, \vec{z}) \mu_{1}(\vec{x}\p, \vec{z}\p) \dif \vec{x} \dif \vec{u}.
\end{align*}
The second identity implies that the resetting is only dependent on the parameter $\vartheta$ and independent of the state and actions $(\vec{x}, \vec{u}, \vec{z})$ to satisfy the ergodicity property.

The parameters $\eta$ and $\lambda$ are the Lagrangian variables associated with Equation~\eqref{eq:reps_kl} and \eqref{eq:reps_norm}, while $V(\vec{x}, \vec{z})$ is the state-value function, which appears naturally in \gls*{REPS} as the Lagrangian function associated with Equation~\eqref{eq:reps_dynamics}. Next, we take the partial derivative of $\mathcal{L}$ with respect to $p(\vec{x}, \vec{u}, \vec{z})$
\begin{align*}
	\frac{\partial \mathcal{L}}{\partial p} =
	\begin{aligned}[t]
		& r(\vec{x}, \vec{u}) - \lambda + (1 - \vartheta) \sum_{\vec{z}\p} \int V(\vec{x}\p, \vec{z}\p) \mu_{1}(\vec{x}\p, \vec{z}\p) \dif \vec{x}\p \\
		& + \vartheta \sum_{\vec{z}\p} \int V(\vec{x}\p, \vec{z}\p) p(\vec{x}\p, \vec{z}\p | \vec{x}, \vec{u}, \vec{z}) \dif \vec{x}\p \\ 
		& - V(\vec{x}, \vec{z}) - \eta \log \frac{p^{*}(\vec{x}, \vec{u}, \vec{z})}{q(\vec{x}, \vec{u}, \vec{z})} - \eta,
	\end{aligned}
\end{align*}
and set it to zero to get the optimal point
\begin{align}
	p^{*}(\vec{x}, \vec{u}, \vec{z}) = q(\vec{x}, \vec{u}, \vec{z}) \exp \left[\frac{A(\vec{x}, \vec{u}, \vec{z}, V) - \lambda - \eta}{\eta} \right], \label{eq:reps_update}
\end{align}
where $A(\vec{x}, \vec{u}, \vec{z}, V)$ is the advantage function given as 
\begin{align}
	\label{eq:reps_adv}
	A(.) & =
	\begin{aligned}[t]
		& r(\vec{x}, \vec{u}) + (1 - \vartheta) \sum_{\vec{z}\p} \int V(\vec{x}\p, \vec{z}\p) \mu_{1}(\vec{x}\p, \vec{z}\p) \dif \vec{x}\p \\
		& \hspace{-0.5cm} + \vartheta \sum_{\vec{z}\p} \int V(\vec{x}\p, \vec{z}\p) p(\vec{x}\p, \vec{z}\p | \vec{x}, \vec{u}, \vec{z}) \dif \vec{x}\p - V(\vec{x}, \vec{z}).
	\end{aligned}
\end{align}

The optimal point $p^{*}(\vec{x}, \vec{u}, \vec{z}) = \mu^{*}(\vec{x}, \vec{z}) \pi^{*}(\vec{u} | \vec{x}, \vec{z})$ has to satisfy the constraint in Equation~\eqref{eq:reps_norm}, which in turn enables us to find the Lagrangian variable $\lambda^{*}$
\begin{align*}
	1           & \! = \! \sum_{\vec{z}} \! \iint \! p^{*}(\vec{x}, \vec{u}, \vec{z}) \dif \vec{x} \dif \vec{u} \\
	1           & \! = \! \sum_{\vec{z}} \! \iint \! q(\vec{x}, \vec{u}, \vec{z}) \exp \left[\displaystyle \frac{A(\vec{x}, \vec{u}, \vec{z}, V) - \lambda^{*} - \eta}{\eta} \right] \! \dif \vec{x} \dif \vec{u} \\
	\lambda^{*} & \! = \! - \eta \! + \! \eta \log \sum_{\vec{z}} \! \iint \! q(\vec{x}, \vec{u}, \vec{z}) \exp \left[\displaystyle \frac{A(\vec{x}, \vec{u}, \vec{z}, V)}{\eta} \right] \! \dif \vec{x} \dif \vec{u}.
\end{align*}
By substituting $\lambda^{*}$ back into $p^{*}(\vec{x}, \vec{u}, \vec{z})$ in Equation~\eqref{eq:reps_update}, we retrieve the normalized density softmax form
\begin{align*}
	p^{*}(\vec{x}, \vec{u}, \vec{z}) = \frac{q(\vec{x}, \vec{u}, \vec{z}) \exp \left[A(\vec{x}, \vec{u}, \vec{z}, V) / \eta \right]}{\sum_{\vec{z}} \iint q(\vec{x}, \vec{u}, \vec{z}) \exp \left[ A(\vec{x}, \vec{u}, \vec{z}, V) / \eta \right] \dif \vec{x} \dif \vec{u}}.
\end{align*}
Now by plugging the solutions $p^{*}$ and $\lambda^{*}$ back into the Lagrangian, we arrive at the dual function $\mathcal{G}$ as a function of the remaining Lagrangian variables $\eta$ and $V$
\begin{equation*}
	\mathcal{G} \! = \! \eta \epsilon + \eta \log \sum_{\vec{z}} \! \iint \! q(\vec{x}, \vec{u}, \vec{z}) \! \exp \left[\frac{A(\vec{x}, \vec{u}, \vec{z}, V)}{\eta} \right] \! \dif \vec{x} \dif \vec{u},
\end{equation*}
where $q(\vec{x}, \vec{u}, \vec{z}) = q(\vec{x}, \vec{u}) q(\vec{z} | \vec{x}, \vec{u})$ and $q(\vec{z} | \vec{x}, \vec{u})$ is the posterior over $\vec{z}$ given $\vec{x}$ and $\vec{u}$. In Section~\ref{sec:sysid}, we derived a forward-backward algorithm for inferring this density, allowing us to compute the expectation over $\vec{z}$. The expectations over $\vec{x}$ and $\vec{u}$ are analytically intractable. Therefore, we approximate them given samples from the reference distribution $q(\vec{x}, \vec{u})$. The multipliers $\eta$ and $V$ are then obtained by numerically minimizing the dual $\mathcal{G}(\eta, V)$
\begin{equation*}
	\argmin_{\eta, V} \quad \mathcal{G}(\eta, V), \quad \quad \st \quad \eta \ge 0,
\end{equation*}
that acts as the upper bound on the primal objective.

\subsection{Modeling Dynamics and State-Value Function}
Up to this point, the derivation of \gls*{Hb-REPS} has been generic. We have made no assumptions on initial distributions $\mu_{1}(\vec{x}, \vec{z})$, the dynamics $p(\vec{x}\p, \vec{z}\p| \vec{x}, \vec{u}, \vec{z})$, or the value function $V(\vec{x}, \vec{z})$. Now, we introduce the piecewise affine-Gaussian dynamics and logistic switching described in Section~\ref{sec:hdbn} and assume these representations to be available in parametric form as a result of a separate learning process. Furthermore, we model the state-value function with piecewise $n$-th degree polynomial functions $V(\vec{x}, \vec{z}) = \vec{\tau}_{\vec{z}}^{\top} \psi_{\vec{z}}(\vec{x})$, where $\psi_{\vec{z}}(\vec{x})$ is the state-feature vector which contains polynomial features of the state $\vec{x}$, and $\vec{\tau}_{\vec{z}}$ is the parameter vector assigned to the different regions. 

Under these assumptions, we can use the available joint density $\mu_{1}(\vec{x}, \vec{z})$ and $p(\vec{x}\p | \vec{x}, \vec{u}, \vec{z})$ to compute the necessary expectations in Equation~\eqref{eq:reps_adv}
\begin{align*}
	& \hspace{-1em} \mathbb{E}_{\vec{x}_{1}, \vec{z}_{1}} \left[V(\vec{x}\p, \vec{z}\p) \right] \! = \! \sum_{\vec{z}\p} \int V(\vec{x}\p, \vec{z}\p) \mu_{1}(\vec{x}\p, \vec{z}\p) \dif \vec{x}\p, \\
	& \hspace{-1em} \mathbb{E}_{\vec{x}\p, \vec{z}\p}\left[V(\vec{x}\p, \vec{z}\p)\right] \! = \! \sum_{\vec{z}\p} \int V(\vec{x}\p, \vec{z}\p) p(\vec{x}\p, \vec{z}\p | \vec{x}, \vec{u}, \vec{z}) \dif \vec{x}\p.
\end{align*}
This computation allows our approach to capture the stochasticity of the dynamics and delivers an estimate of the advantage function $A(\vec{x}, \vec{u}, \vec{z}, V)$ instead of the \gls*{TD} error in the general REPS framework \cite{peters2010relative}. Ultimately, this leads to better estimates of the expected discounted future returns captured by $V$.

Practically, these integrals can be either naively approximated by applying Monte Carlo integration \cite{robert1999monte} or, more efficiently, by recognizing the structure of the integrand $V(\vec{x}\p, \vec{z}\p)$ and using Gauss-Hermite cubature rules for exact integration over polynomial functions \cite{sarkka2013bayesian}.

\subsection{Maximum-A-Posteriori Policy Improvement}
\label{sec:policy}
A significant advantage of our model-based reinforcement learning approach becomes evident when considering the policy improvement step in the \gls*{REPS} framework. The policy update is incorporated into the optimality condition of the stationary state-action distribution $p(\vec{x}, \vec{u}, \vec{z}) = \pi(\vec{u} | \vec{x}, \vec{z}) \mu(\vec{x}, \vec{z})$ in Equation~\eqref{eq:reps_update}. As a consequence, updating the mixture policies $\pi(\vec{u} | \vec{x}, \vec{z})$ requires the computation of state probabilities $\mu(\vec{x}, \vec{z})$, which in turn require knowledge of the dynamics model. This issue is circumvented in other model-free realizations of \gls*{REPS} by introducing a crude approximation to enable a model-free policy update nonetheless. For example, in \cite{deisenroth2013survey}, the authors postulate that the new state distribution $\mu(\vec{x}, \vec{z})$ is usually \emph{close enough} to the old distribution $q(\vec{x}, \vec{z})$, thus allowing the ratio $q(\vec{x}, \vec{z}) / \mu(\vec{x}, \vec{z})$ to be ignored when a weighted maximum-likelihood fit of the actions $\vec{u}$ is performed to update $\pi$. 

While the assumption of \emph{closeness} may be practical and empowers many successful variants of \gls*{REPS}, it is crucial to be aware of its technical ramifications, as it undermines the primary motivation of a relative entropy bound on the state-action distribution in Equation~\eqref{eq:reps_kl}. This aspect is unique in the \gls*{REPS} framework when compared to other state-of-the-art approximate policy iteration algorithms \cite{schulman2015trust, schulman2017proximal, haarnoja2018soft}, that optimize a similar objective, albeit with a relaxed bound that only limits the change of the action distribution $\pi$.

In contrast, our algorithm uses the surrogate hybrid dynamics and updates the policy $\pi(\vec{u} | \vec{x}, \vec{z})$ with the correct weighting. The optimality condition in Equation~\eqref{eq:reps_update} is satisfied by computing a weighted maximum a posteriori estimate of the parameters $\vec{\theta}$ of the state-action distribution $p(\vec{x}, \vec{u}, \vec{z} | \vec{\theta})$, thus implicitly updating $\pi(\vec{u} | \vec{x}, \vec{z})$. This procedure is equivalent to a modified Baum-Welch expectation-maximization algorithm that learns the parameters of a closed-loop \gls*{rARHMM}, as derived in Section~\ref{sec:sysid}. The difference is that the \gls*{EM} objective in Equation~\eqref{eq:log_prob} has to be augmented with the importance weights from Equation~\eqref{eq:reps_update}
\begin{equation*}
	\argmax_{\vec{\theta}} ~ \log \prod_{n=1}^{N} \sum_{\vec{z}^{n}} \vec{w}^{n} p(\mat{X}^{n}, \mat{U}^{n}, \mat{Z}^{n} | \vec{\theta}) p(\vec{\theta}),
\end{equation*}
where $(\mat{X}^{n}, \mat{U}^{n})$ are state-action trajectories collected via interaction with the environment and $\vec{w}^{n}$ are the associated weights resulting from Equation~\eqref{eq:reps_update}
\begin{equation*}
	\vec{w}^{n} = \exp \left[ A(\mat{X}^{n}, \mat{U}^{n}, \mat{Z}^{n}, V) / \eta \right].
\end{equation*}
This augmentation leads to weighted M- and EB-steps while the E-step is not altered. 

Note that during the policy improvement step, we can either assume an a priori estimate of the open-loop dynamics $p(\vec{x}\p, \vec{z}\p | \vec{x}, \vec{u}, \vec{z})$ and only update the control parameters corresponding to the conditional $\pi(\vec{u} | \vec{x}, \vec{z})$, or we can iteratively update $p(\vec{x}\p, \vec{z}\p | \vec{x}, \vec{u}, \vec{z})$ as more data becomes available. A compact sketch of the overall optimization process is available in Algorithm~\ref{alg:reps_algo}.

\removealgorithmerror
\SetEndCharOfAlgoLine{}
\RestyleAlgo{boxruled}
\begin{algorithm}[t!]
	\linespread{1.25}\selectfont
	\SetKwInput{Input}{input}
	\SetKwInput{Output}{output}
	\SetKwInput{Initialize}{initialize}
	\SetKwRepeat{Do}{do}{while}
	
	\Input{$p(\vec{x}\p, \vec{z}\p | \vec{x}, \vec{u}, \vec{z})$}
	\Initialize{$q(\vec{u} | \vec{x} , \vec{z}), \vec{\tau}_{\vec{z}}, \eta$}
	
	\While{$J$ \normalfont not converged}
	{
		\tcp{Sample interactions}
		$(\mat{X}, \mat{U}) \leftarrow$ \bf{Environment}$(q)$ \\
		
		\tcp{Policy evaluation}
		$\eta^{*}, \vec{\tau}^{*}_{\vec{z}}, \vec{w}^{*} \leftarrow$ \text{\bf Minimize} $\mathcal{G}(\mat{X}, \mat{U}, p, \eta, \vec{\tau}_{\vec{z}}, \epsilon)$
		
		\tcp{Policy improvement}
		$\pi^{*} \leftarrow$ \text{\bf BaumWelch}$(\mat{X}, \mat{U}, p, \vec{w}^{*})$
		
		\tcp{Update parameters}
		$q, \vec{\tau}_{\vec{z}}, \eta \leftarrow \pi^{*}, \vec{\tau}^{*}_{\vec{z}}, \eta^{*}$
	}
	\Output{$\pi^{*}(\vec{u} | \vec{x}, \vec{z})$}
	
	\linespread{1.0}\selectfont
	\caption{Model-Based Relative Entropy Policy Search via Hybrid Models}
	\label{alg:reps_algo}
\end{algorithm}

\section{Empirical Evaluation}
\label{sec:eval}
In this section, we benchmark different aspects of our approach to system modeling and control synthesis via hybrid models. In the following,
\begin{itemize}
	\item we assess the predictive performance of \glspl*{rARHMM} at open-loop system identification of nonlinear systems and validate our choice of hybrid surrogate models as a suitable representation.
	\item we test the ability of \glspl*{rARHMM} to approximate and decompose expert nonlinear controllers in a closed-loop behavioral cloning scenario.
	\item we deploy \glspl*{rARHMM} in the proposed hierarchical RL algorithm \gls*{Hb-REPS} to solve the infinite horizon stochastic control objective and optimize piecewise polynomial controllers and value functions.
\end{itemize}

\begin{figure*}[t!]
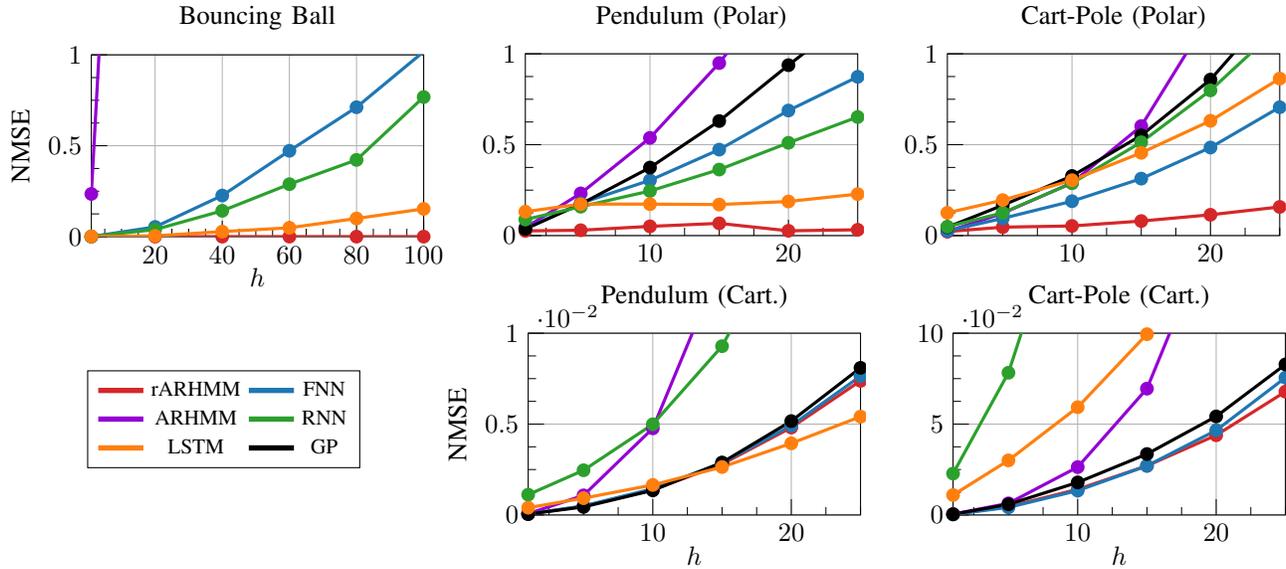

	\begin{adjustbox}{valign=t,minipage={.33\textwidth}}
		\input{figures/bouncing_nmse.tex}%
	\end{adjustbox}
	\begin{adjustbox}{valign=t,minipage={.33\textwidth}}
		\hspace{0.2cm}
		\input{figures/joint_pendulum_nmse.tex}%
	\end{adjustbox}
	\begin{adjustbox}{valign=t,minipage={.33\textwidth}}
		\input{figures/joint_cartpole_nmse.tex}%
	\end{adjustbox}
	\begin{adjustbox}{valign=t,minipage={.33\textwidth}}
		\vspace{1.cm}
		\hspace{1.cm}
		\input{figures/identification_legend.tex}%
	\end{adjustbox}
	\begin{adjustbox}{valign=t,minipage={.33\textwidth}}
		\vspace{-0.25cm}
		\hspace{-0.25cm}
		\input{figures/cart_pendulum_nmse.tex}%
	\end{adjustbox}
	\begin{adjustbox}{valign=t,minipage={.33\textwidth}}
		\vspace{-0.25cm}
		\hspace{0.05cm}
		\input{figures/cart_cartpole_nmse.tex}%
	\end{adjustbox}
	\vspace{-0.15cm}
	\caption{System identification: The $h$-step \gls*{NMSE} of \glspl*{rARHMM} compared to other models. Evaluation is averaged over 24 data splits. Benchmarking on three dynamical systems, a bouncing ball, a pendulum, and a cart-pole. \glspl*{rARHMM} exhibit the most consistent approximation capabilities. Figure reproduced from \cite{abdulsamad2020hierarchical}.}
	\label{fig:sysid}
	\vspace{0.0cm}
\end{figure*}

\begin{table*}[htb]
	\centering
	\begin{tabular}{c c c c c c}
		\toprule
		& Bouncing Ball & Pendulum (Polar) & Pendulum (Cartesian) & Cart-Pole (Polar) & Cart-Pole (Cartesian) \\
		\midrule
		ARHMM  & 22 (2)        & 180 (9)          & 130 (5)              & 287 (7)           & 275 (5) \\
		\midrule
		rARHMM & 86 (2)        & 468 (9)          & 582 (9)              & 575 (7)           & 711 (7) \\
		\midrule
		FNN    & 1250 (32)     & 546 (64)         & 1315 (32)            & 1380 (32)         & 1445 (32) \\
		\midrule
		RNN    & 12866 (64)    & 50306 (128)      & 3427 (32)            & 50820 (128)       & 51077 (128) \\
		\midrule
		LSTM   & 200450 (128)  & 51074 (64)       & 51395 (64)           & 201732 (128)      & 202373 (128) \\
		\bottomrule
	\end{tabular}
	\caption{System identification: Qualitative comparison of model complexity for the best-performing representations in Figure~\ref{fig:sysid}. The values reflect the total number of parameters of each model. The values in parentheses represent the hidden layer sizes $S$ of the neural models and the number of discrete components $K$ for the (r)ARHMM, respectively.}
	\label{table:complexity}
	\vspace{-0.2cm}
\end{table*}

\subsection{Piecewise Open-Loop System Identification}
We start by empirically benchmarking the open-loop learned \glspl*{rARHMM} and their ability to approximate nonlinear dynamics. We compare to popular black-box models in a \emph{long-horizon} and \emph{limited-data} setting.

This evaluation focuses on \glspl*{rARHMM} with exogenous inputs. We learn the dynamics of three simulated deterministic systems; a bouncing ball, an actuation-constrained pendulum, and a cart-pole system. We compare the predictive time forecasting accuracy of \glspl*{rARHMM} to classical non-recurrent \glspl*{ARHMM}~\footnote{\glspl*{ARHMM} closely resemble \glspl*{rARHMM}. However, the transitions probability in Equation~\ref{eq:trans_probs} does not depend on the continuous state or action.}\cite{fox2009bayesian}, \glspl*{FNN}, \glspl*{GP}~\footnote{With an RBF kernel and hyperparameter optimization.}, \glspl*{LSTM} \cite{hochreiter1997long}, and \glspl*{RNN}. During the evaluation, we collected segregated training and test datasets. The training dataset is randomly split into 24 groups, each used to train different instances of all models. These instances are then tested on the test dataset. During evaluation, we sweep the test trajectories stepwise and predict the given horizon.

All neural models have two hidden layers, which we test for different layer sizes, $S \in \{16, 32, 64, 128, 256, \allowbreak 512\}$ for \glspl*{FNN}, $S \in \{16, 32, 64, 128, \allowbreak 256\}$ for \glspl*{RNN}, and $S \in \{16, 32, 64, 128\}$ for \glspl*{LSTM}. In the case of (r)ARHMMs, we vary the number of components $K$, dependent on the task. As a metric, we evaluate the forecast \gls*{NMSE} for a range of horizons averaged over the 24 data splits. We report the result corresponding to the best choice of $S$ and $K$. Finally, in Table~\ref{table:complexity}, we qualitatively compare the complexity of all representations in terms of their total number of parameters.

\begin{figure*}[!t]
	\begin{minipage}{0.55\textwidth}
		\begin{minipage}[t]{0.33\textwidth}%
			\input{figures/pendulum_openloop.tex}%
		\end{minipage}\hspace*{1.5cm}%
		\begin{minipage}[t]{0.33\textwidth}%
			\input{figures/sac_pendulum_imitation_phase.tex}%
		\end{minipage}
	\end{minipage}
	\begin{minipage}{0.45\textwidth}
		\vspace{-0.75cm}
		\caption{Behavioral cloning: Phase space of the pendulum. The identified unforced dynamics are on the left (blue). The learned model qualitatively captures the phase portrait. On the right (red) are the closed-loop dynamics. The learned stationary hybrid policy with five regions successfully imitates a global nonlinear \gls*{SAC} controller to stabilize the system around the origin. Figure reproduced from \cite{abdulsamad2020hierarchical}.}
		\label{fig:pendulum_phase}
	\end{minipage}
	\vspace{-0.35cm}
\end{figure*}

\begin{figure*}[htb]
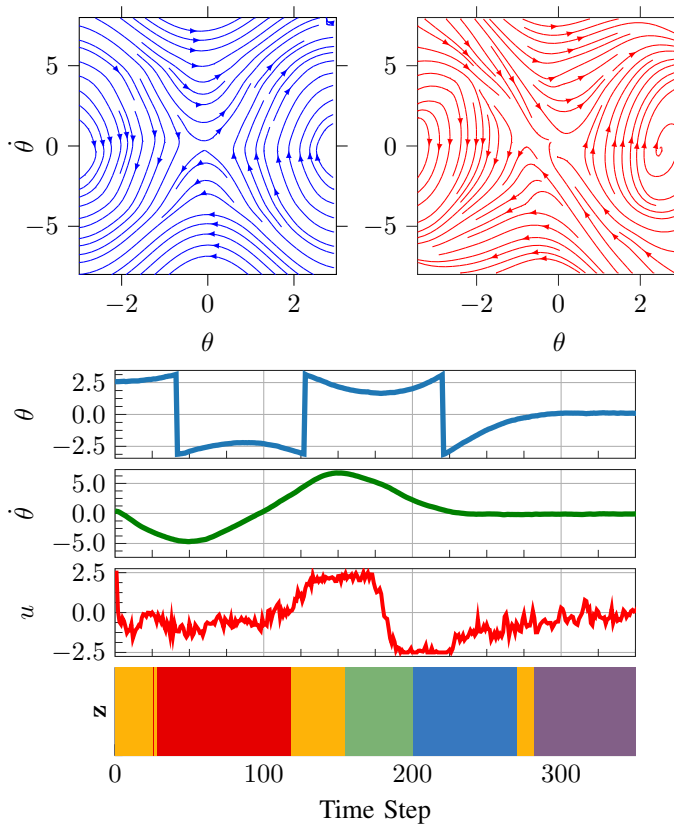

	\begin{minipage}{0.45\textwidth}
		\input{figures/gps_pendulum_imitation_example.tex}
	\end{minipage}\hspace{1.0cm}
	\begin{minipage}{0.45\textwidth}
		\input{figures/sac_cartpole_imitation_example.tex}
	\end{minipage}
	\vspace{-0.5cm}
	\caption{Behavioral cloning: Sample trajectories from the learned hybrid policies on the pendulum (left) and cart-pole (right). Both hybrid controllers are able to consistently solve the tasks while relying on simple local representations of the feedback controllers. The colors indicate the active dynamics and control regimes over time. Figure reproduced from \cite{abdulsamad2020hierarchical}.}
	\label{fig:imitation_trajectories}
	\vspace{-0.55cm}
\end{figure*}

\subsubsection{Bouncing Ball} This example is a canonical instance of a dual-regime hybrid system due to the hard velocity switch at the moment of impact. We simulate the dynamics with a frequency of \SI{20}{\hertz} and collect 25 training trajectories with different initial heights and velocities, each \SI{30}{\second} long. This dataset is split 24 folds with ten trajectories, $10\times150$ data points, in each subset. The test dataset consists of 5 trajectories, each \SI{30}{\second} long. We evaluate the \gls*{NMSE} for horizons $h =\{1, 20, 40, 60, 80\}$ time steps. We did not evaluate a GP model in this setting due to the long prediction horizons that led to a very high computational burden. The (r)ARHMMs are tested for $K=2$. The logit link function of an \gls*{rARHMM} is parameterized by a neural net with one hidden layer containing 16 neurons. The results in Figure~\ref{fig:sysid} show that the \gls*{rARHMM} approximates the dynamics well and outperforms both \glspl*{ARHMM} and the neural models.

\subsubsection{Pendulum and Cart-Pole} These systems are classical benchmarks from the nonlinear control literature. Here we consider two different observation types, one in the wrapped polar space, where the angle space $\theta \in [-\pi, \, \pi]$ includes a sharp discontinuity, and a second model with smooth observations parameterized with the Cartesian trigonometric features $\{ \cos(\theta), \sin(\theta) \}$. Both dynamics are simulated with a frequency of \SI{100}{\hertz}. We collect 25 training trajectories starting from different initial conditions and apply random uniform explorative actions. Each trajectory is \SI{2.5}{\second} long. The 24 splits consist of 10 trajectories each, $10\times250$ data points. The test dataset consists of 5 trajectories, each \SI{2.5}{\second} long. Forecasting accuracy is evaluated for horizons $h=\{1, 5, 10, 15, 20, 25\}$. The (r)ARHMMs are tested for $K=\{3, 5, 7, 9\}$ on both tasks. The logit link function of the \gls*{rARHMM} is parameterized by a neural net with one hidden layer containing 24 neurons. As shown in Figure~\ref{fig:sysid}, the forecast evaluation provides empirical evidence for the representation power of \glspl*{rARHMM} in both smooth and discontinuous state spaces. \glspl*{FNN} and \glspl*{GP} perform well in the smooth Cartesian observation space and struggle in the discontinuous space, similar to \glspl*{RNN} and \glspl*{LSTM}. Moreover, in Table~\ref{table:complexity}, it is clear that \glspl*{rARHMM} reach comparable predictive performance to state-of-the-art models with a fraction of the parametric complexity.

\subsection{Piecewise Closed-Loop Behavioral Cloning}
We want to analyze the closed-loop \gls*{rARHMM} with endogenous inputs as a behavioral cloning framework. The task is to reproduce the closed-loop behavior of expert policies on challenging nonlinear systems. For this purpose, we train two feedback experts on the pendulum and cart-pole. The two environments are simulated at \SI{50}{\hertz} and are influenced by static Gaussian noise with a standard deviation $\sigma = \num{1e-2}$. The experts are two-layer neural network policies with 4545 parameters (pendulum) and 17537 parameters (cart-pole), optimized with the \gls*{SAC} algorithm \cite{haarnoja2018soft}. 

For cloning, we construct two 5-regime \glspl*{rARHMM} with piecewise polynomial policies of the third order. The hybrid controllers have a total number of parameters of 100 (pendulum) and 280 (cart-pole). Learning is realized on a dataset of 25 expert trajectories, each \SI{5}{\second} long, for each environment and using the \gls*{EM} technique from Section~\ref{sec:sysid}. The decomposed controllers complete the task of swinging up and stabilizing both systems with over 95$\%$ success rate. Figure~\ref{fig:pendulum_phase} shows the phase portraits of the unforced dynamics and closed-loop control identified during cloning. Figure~\ref{fig:imitation_trajectories} depicts sampled trajectories of the hybrid policies highlighting the switching behavior.

\begin{figure*}[!t]
	\begin{minipage}{0.55\textwidth}
		\begin{minipage}[t]{0.33\textwidth}%
			\input{figures/pole_openloop.tex}%
		\end{minipage}\hspace*{1.5cm}%
		\begin{minipage}[t]{0.33\textwidth}%
			\vspace{-4.5cm}
			\includeinkscape[width=115pt, svgpath=figures/]{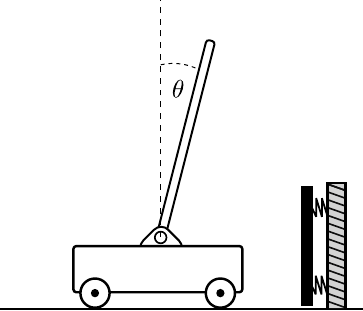}
		\end{minipage}
	\end{minipage}
	\vspace{-.25cm}
	\begin{minipage}{0.45\textwidth}
		\vspace{-0.5cm}
		\caption{A cart-pole system with an elastic wall constraint: The cart-pole dynamics are linearized around the upright position, and a spring system models the wall. The phase portrait of the unforced angular dynamics is depicted on the left (blue). The aim is to stabilize the pole around the origin.}
		\label{fig:pole_phase}
	\end{minipage}
	\vspace{-0.1cm}
\end{figure*}

\begin{figure*}[htb]
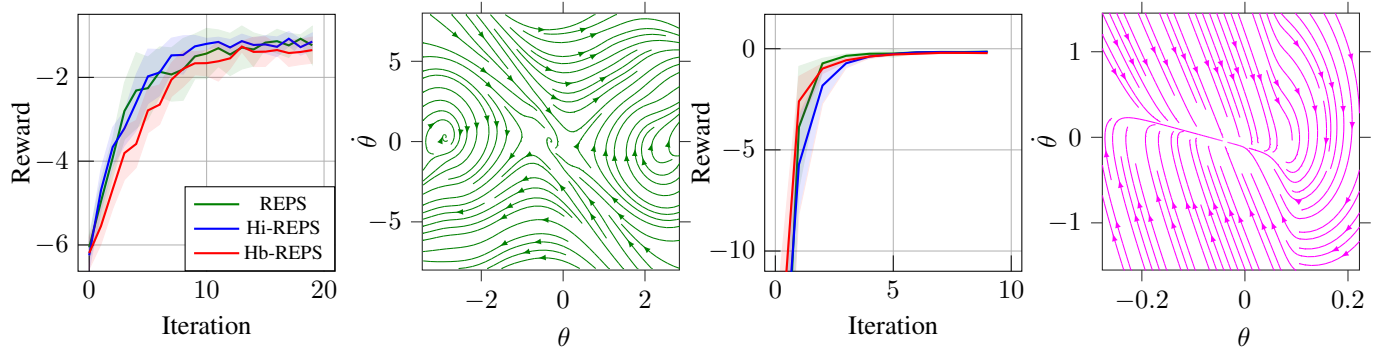

	\begin{minipage}{0.22\textwidth}
		\input{figures/pendulum_reps.tex}
	\end{minipage}\hspace{.5cm}
	\begin{minipage}{0.22\textwidth}
		\input{figures/hbreps_pendulum_rl_phase.tex} %
	\end{minipage}\hspace{.5cm}
	\begin{minipage}{0.22\textwidth}
		\input{figures/pole_reps.tex}
	\end{minipage}\hspace{.5cm}
	\begin{minipage}{0.22\textwidth}
		\input{figures/hbreps_pole_rl_phase.tex} %
	\end{minipage}
	\vspace{-0.5cm}
	\caption{Reinforcement learning: \gls*{REPS}, Hi-REPS, and \gls*{Hb-REPS} evaluated on the pendulum swing-up (left) and cart-pole stabilization (right) tasks. By inspecting the learning curves, mean reward with two standard deviations, we conclude that all algorithms perform equally well in terms of the transient and final performance. However, \gls*{Hb-REPS} relies on simpler piecewise polynomial models of the policy and value function, while Hi-REPS and \gls*{REPS} use nonlinear Fourier basis functions. The phase portraits depict the closed-loop behavior achieved by \gls*{Hb-REPS}.}
	\label{fig:hbreps_rl}
	\vspace{-0.55cm}
\end{figure*}

\subsection{Nonlinear Control Synthesis via Hybrid Models}
Finally, we evaluate the performance of the hybrid policy search algorithm \gls*{Hb-REPS} on two nonlinear stochastic dynamical systems: an actuation-constrained pendulum swing-up and a cart-pole stabilization task. We make no claim to the absolute \emph{sample} efficiency of our \gls*{RL} approach when compared to state-of-the-art RL algorithms. Instead, we aim to provide empirical support for the premise that structured representations that rely on compact piecewise parametric forms can provide an alternative to black-box function approximators with comparable overall performance. 

Therefore, we compare the performance of \gls*{Hb-REPS} to two baselines. The first is a \emph{vanilla} version of \gls*{REPS} that does not maintain any hierarchical structure and uses nonlinear function approximators with \glspl*{RFF} \cite{rahimi2008random} to represent both policy and value function. The second baseline assumes a hierarchical policy structure and a nonlinear value function with Fourier features. This baseline is somewhat akin to what is implemented in \cite{daniel2016probabilistic}, albeit with a hierarchy based on state abstraction rather than time. We will refer to this algorithm as hierarchical REPS (Hi-REPS). We assume an offline learning phase in which the hybrid models are learned from pre-collected data.

\subsubsection{Pendulum Swing-up} In this experiment, the actuation-constrained pendulum is simulated at \SI{50}{\hertz} and perturbed by Gaussian noise with a standard deviation $\sigma = \num{1e-2}$. The \gls*{REPS} agent relies on a policy and value function with 50 and 75 Fourier basis functions, respectively. Hi-REPS assumes a similar form of the value function but with a piecewise third-order polynomial policy over five partitions. \gls*{Hb-REPS} represents both policy and value function with piecewise third-order polynomials over five partitions. Empirical results in Figure~\ref{fig:hbreps_rl} (left half) feature comparable learning performance of all algorithms over ten random seeds. Every iteration involves 5000 interactions with the environment. We provide a phase portrait of the closed-loop behavior for a qualitative assessment of the final stationary hybrid policy.

\subsubsection{Cart-pole Stabilization} This evaluation features a cart-pole constrained by an elastic wall modeled by a spring. The dynamics are linearized around the upright position. The environment is simulated at \SI{100}{\hertz} and perturbed by Gaussian noise with a standard deviation $\sigma = \num{1e-4}$. The REPS policy and value function both use 25 random Fourier basis functions. Hi-REPS adopts the same value function structure with a two-partition piecewise affine policy. \gls*{Hb-REPS} also assumes a two-partition piecewise affine policy and second-order value function. Figure~\ref{fig:hbreps_rl} (right half) depicts comparable learning performance over ten random seeds. Every iteration involves 2500 interactions with the environment.

\section{Discussion}
\label{sec:conclusion}
We presented a general framework for data-driven nonlinear system identification and stochastic control based on the structured representation of hybrid surrogate models. To introduce the hybrid structure, we proposed replacing commonly used piecewise affine auto-regressive models with probabilistic hybrid dynamic Bayesian networks, as they offer a range of advantages in data-driven scenarios. Furthermore, we presented a novel reinforcement learning algorithm that leverages the learned hybrid models to synthesize piecewise polynomial feedback controllers for nonlinear systems. 

Our hybrid-model-infused reinforcement learning approach is able to reach comparable performance on control tasks with a significant reduction in the complexity of functional representation. Furthermore, in contrast to deterministic hybrid model predictive control, our approach solves the infinite-horizon stochastic optimal control problem by approximating the global value function and lifts the requirement for polyhedral partitioning.

While initial empirical results are encouraging, the application of this work is limited to low-dimensional dynamical systems. Although a viable alternative to expensive mixed-integer optimization, the inference techniques used in this paper still present a bottleneck in the face of scalability to higher dimensions. While our \gls*{MAP} approach significantly improves the quality of expectation-maximization solutions, it nevertheless struggles in more challenging environments.

A possible course of action is to investigate Bayesian nonparametric extensions of hybrid dynamic Bayesian networks based on non-conjugate variational inference. Fully Bayesian methods tend to improve learning in large structured models significantly. Another potential avenue of research is to improve the hybrid reinforcement learning framework by considering the control-as-inference paradigm. Such approaches may offer ways of integrating the Bayesian structure of the models into the control optimization and constructing an uncertainty-aware approach that is better equipped to deal with the exploration-exploitation dilemma.

\vspace{-.25cm}
\appendices

\renewcommand{\thesectiondis}[2]{\Alph{section}:}

\section{Exponential Family}
\label{app:expo}
Our work focuses on random variables with probability density functions belonging to the exponential family. The unified minimal parameterization of this class of distributions lends itself for convenient and efficient posterior computation when paired with conjugate priors.

We assume the natural form for a probability density of a random variable $\vec{x}$
\begin{equation*}
	f(\vec{x} | \vec{\eta}) = h(\vec{x}) \exp \left[\vec{\eta} \cdot \vec{t}(\vec{x}) - a(\vec{\eta}) \right],
\end{equation*}
where $h(\vec{x})$ is the base measure, $\vec{\eta}$ are the natural parameters, $\vec{t}(\vec{x})$ are the sufficient statistics and $a(\vec{\eta})$ is the log-partition function, or log-normalizer. Following the same notation, a conjugate prior $g(\vec{\eta} | \vec{\lambda})$ to the likelihood $f(\vec{x} | \vec{\eta})$ has the form
\begin{equation*}
	g(\vec{\eta} | \vec{\lambda}) = h(\vec{\eta}) \exp \left[\vec{\lambda} \cdot \vec{t}(\vec{\eta}) - a(\vec{\lambda}) \right],
\end{equation*}
with prior sufficient statistics $\vec{t}(\vec{\eta}) = [\vec{\eta}, \, - a(\vec{\eta})]^{\top}$ and hyperparameters $\vec{\lambda} = [\vec{\alpha}, \, \vec{\beta}]^{\top}$. By applying Bayes' rule, we can directly infer the posterior $q(\vec{\eta} | \vec{x})$
\begin{align*}
	q(\vec{\eta} | \vec{x}) & \propto f(\vec{x} | \vec{\eta}) g(\vec{\eta} | \vec{\lambda})                                            \\
	& \propto \exp \left[\vec{\rho}(\vec{x}, \vec{\lambda}) \cdot \vec{t}(\vec{\eta}) - a(\vec{\rho}) \right],
\end{align*}
where the posterior natural parameters $\vec{\rho}(\vec{x}, \vec{\lambda})$ are a function of the likelihood sufficient statistics $\vec{t}(\vec{x})$ and prior hyperparameters $[\vec{\alpha}, \, \vec{\beta}]$
\begin{equation*}
	\vec{\rho}(\vec{x}, \vec{\lambda}) = \left[\vec{\alpha} + \vec{t}(\vec{x}), ~ \vec{\beta} + \vec{1} \right]^{\top}.
\end{equation*}
The structure of the resulting posterior reveals a simple recipe for data-driven inference. By moving into the natural space, the posterior parameters are computed by combining the prior hyperparameters with the likelihood sufficient statistics and log-partition function. By definition, every exponential family distribution has a minimal natural parameterization that leads to a unique decomposition of these quantities \cite{wainwright2008graphical}.

\section{Conjugate Posteriors}
\label{app:posteriors}
We present an outline of all \gls*{M-step} updates. We use an adapted form of the exponential natural parameterization, as it offers a clear methodology for deriving and implementing such updates for all relevant distributions.

\subsection{Categorical with Dirichlet Prior}
A weighted categorical likelihood over a one-hot random variable $\vec{z}$ with size $K$ has the form
\begin{align*}
	p(\mat{Z} | \vec{\varphi}) & = \prod_{n=1}^{N} \Cat(\vec{z}_{n} | \vec{\varphi})^{w_{n}} \\
	& \propto
	\exp \left\{
	\begin{bmatrix}
		\log \varphi{1} \\
		\vdots       \\
		\log \varphi{K}
	\end{bmatrix}
	\cdot
	\begin{bmatrix}
		\sum_{n=1}^{N} w_{n,1} \\
		\vdots                 \\
		\sum_{n=1}^{N} w_{n,K}
	\end{bmatrix}
	\right\},
\end{align*}
where $w_{nk}$ are the importance weights for each category $K$. The conjugate prior is a Dirichlet $p(\vec{\varphi})$ distribution
\begin{align*}
	p(\vec{\varphi}) & = \mathrm{Dir}(\vec{\varphi} | \vec{\tau}_{0}) \\
	& \propto
	\exp \left\{
	\begin{bmatrix}
		\tau_{0,1} - 1 \\
		\vdots         \\
		\tau_{0,K} - 1
	\end{bmatrix}
	\cdot
	\begin{bmatrix}
		\log \varphi_{1} \\
		\vdots       \\
		\log \varphi_{K}
	\end{bmatrix}
	\right\},
\end{align*}

The posterior $q(\vec{\varphi})$ is likewise a Dirichlet distribution
\begin{align*}
	q(\vec{\varphi}) & = \mathrm{Dir}(\vec{\varphi} | \vec{\tau}) \\
	& \propto
	\exp \left\{
	\begin{bmatrix}
		\tau_{0,1} - 1 + \sum_{n=1}^{N} w_{n,1} \\
		\vdots                                  \\
		\tau_{0,K} - 1 + \sum_{n=1}^{N} w_{n,K}
	\end{bmatrix}
	\cdot
	\begin{bmatrix}
		\log \varphi_{1} \\
		\vdots       \\
		\log \varphi_{K}
	\end{bmatrix}
	\right\}.
\end{align*}

The maximization step requires computing the mode categorical weights. For a Dirichlet distribution the mode weights are $\hvec{\varphi} = (\vec{\tau} - 1) / (\sum_{k=1}^{K} \tau_{k} - K)$ with $\tau_{k} > 1$. The parameter vector $\vec{\tau}$ is given by
\begin{equation*}
	\tau_{k} = \tau_{0, k} + \sum_{n=1}^{N} w_{n,k} \quad \forall k \in [1, \, K].
\end{equation*}

\subsection*{Gaussian with Normal-Wishart Prior}
A weighted Gaussian likelihood over a random variable $\vec{x} \in \mathds{R}^{d}$ has the following precision-based parameterization
\begin{align*}
	p(\mat{X} | \vec{\mu}, \mat{\Lambda}) & = \prod_{n=1}^{N} \N(\vec{x}_{n} | \vec{\mu}, \mat{\Lambda})^{w_{n}} \\
	& \propto
	\exp \left\{
	\begin{bmatrix}
		\mat{\Lambda} \vec{\mu}                  \\[0.5em]
		\vec{\mu}^{\top} \mat{\Lambda} \vec{\mu} \\[0.5em]
		\mat{\Lambda}                            \\[0.5em]
		\log |\mat{\Lambda}|
	\end{bmatrix}
	\! \cdot \!
	\begin{bmatrix}
		\sum_{n=1}^{N} w_{n} \vec{x}_{n}                                 \\[0.5em]
		-\frac{1}{2} \sum_{n=1}^{N} w_{n}                                \\[0.5em]
		-\frac{1}{2} \sum_{n=1}^{N} w_{n} \vec{x}_{n} \vec{x}_{n}^{\top} \\[0.5em]
		\frac{1}{2} \sum_{n=1}^{N} w_{n}
	\end{bmatrix}
	\right\},
\end{align*}
where $w_{n}$ are the importance weights. The conjugate prior $p(\vec{\mu}, \mat{\Lambda})$ is a normal-Wishart distribution with zero mean
\begin{align*}
	p(\vec{\mu}, \mat{\Lambda}) & = \N(\vec{\mu} | \vec{0}, \kappa_{0} \mat{\Lambda}) \W(\mat{\Lambda} | \mat{\Psi}_{0}, \nu_{0}) \\
	& \propto
	\exp \left\{
	\begin{bmatrix}
		\vec{0}                          \\[0.5em]
		-\frac{1}{2} \kappa_{0}          \\[0.5em]
		-\frac{1}{2} \mat{\Psi}_{0}^{-1} \\[0.5em]
		\frac{1}{2} (\nu_{0} - d)
	\end{bmatrix}
	\! \cdot \!
	\begin{bmatrix}
		\mat{\Lambda} \vec{\mu}                  \\[0.5em]
		\vec{\mu}^{\top} \mat{\Lambda} \vec{\mu} \\[0.5em]
		\mat{\Lambda}                            \\[0.5em]
		\log |\mat{\Lambda}|
	\end{bmatrix}
	\right\}.
\end{align*}
The resulting posterior $q(\vec{\mu}, \mat{\Lambda})$ is also a normal-Wishart
\begin{align*}
	q(\vec{\mu}, \mat{\Lambda}) & \! = \! \N(\vec{\mu} | \vec{m}, \kappa \mat{\Lambda}) \W(\mat{\Lambda} | \mat{\Psi}, \nu) \\
	\!                          & \! \propto
	\exp \left\{ \!
	\begin{bmatrix}
		\sum_{n=1}^{N} w_{n} \vec{x}_{n}                                                         \\[0.5em]
		-\frac{1}{2} (\kappa_{0} + \sum_{n=1}^{N} w_{n})                                         \\[0.5em]
		-\frac{1}{2} (\mat{\Psi}_{0}^{-1} + \sum_{n=1}^{N} w_{n} \vec{x}_{n} \vec{x}_{n}^{\top}) \\[0.5em]
		\frac{1}{2} (\nu_{0} - d + \sum_{n=1}^{N} w_{n})
	\end{bmatrix}
	\! \cdot \!
	\begin{bmatrix}
		\mat{\Lambda} \vec{\mu}                  \\[0.5em]
		\vec{\mu}^{\top} \mat{\Lambda} \vec{\mu} \\[0.5em]
		\mat{\Lambda}                            \\[0.5em]
		\log |\mat{\Lambda}|
	\end{bmatrix}
	\! \right\}.
\end{align*}
The vector and matrix modes of a normal-Wishart distribution are $\hvec{\mu} = \vec{m}$ and $\hmat{\Lambda} = (\nu - d) \mat{\Psi}$, respectively. The posterior parameters needed to determine the modes are
\begin{gather*}
	\kappa = \kappa_{0} + \sum_{n=1}^{N} w_{n}, \vec{m} = 1 / \kappa \sum_{n=1}^{N} w_{n} \vec{x}_{n}, \\
	\nu = \nu_{0} + \sum_{n=1}^{N} w_{n}, \mat{\Psi} = (\mat{\Psi}_{0}^{-1} + \sum_{n=1}^{N} w_{n} \vec{x}_{n} \vec{x}_{n}^{\top} - \kappa \, \vec{m} \, \vec{m}^{\top})^{-1}.
\end{gather*}

\subsection*{Linear-Gaussian with Matrix-Normal-Wishart Prior}
A weighted linear-Gaussian likelihood takes a random variable $\vec{x} \in \mathds{R}^{d}$ and returns a random variable $\vec{y} \in \mathds{R}^{m}$ according to a linear mapping $\mat{A}: \mathds{R}^{d} \to \mathds{R}^{m}$
\begin{align*}
	p(\mat{Y} | \mat{X}, \mat{A}, \mat{V}) & = \prod_{n=1}^{N} \N(\vec{y}_{n} | \vec{x}_{n}, \mat{A}, \mat{V})^{w_{n}} \\
	& \propto
	\exp \left\{
	\begin{bmatrix}
		\mat{V} \mat{A}                \\[0.5em]
		\mat{A}^{\top} \mat{V} \mat{A} \\[0.5em]
		\mat{V}                        \\[0.5em]
		\log |\mat{V}|
	\end{bmatrix}
	\cdot
	\begin{bmatrix}
		\mat{Y} \mat{W} \mat{X}^{\top}              \\[0.5em]
		-\frac{1}{2} \mat{X} \mat{W} \mat{X}^{\top} \\[0.5em]
		-\frac{1}{2} \mat{Y} \mat{W} \mat{Y}^{\top} \\[0.5em]
		\frac{1}{2} \sum_{n=1}^{N} w_{n}
	\end{bmatrix}
	\right\},
\end{align*}
where $w_{n}$ are the weights and $\mat{W} = \mathrm{diag}(w_{n})$ is the diagonal weight matrix. The data matrices $\mat{X}$ and $\mat{Y}$ are of size $d \times N$ and $m \times N$, respectively. The conjugate prior $p(\mat{A}, \mat{V})$ is a matrix-normal-Wishart with zero mean
\begin{align*}
	p(\mat{A}, \mat{V}) & = \N(\mat{A} | \mat{0}, \mat{V}, \mat{K}_{0}) \W(\mat{V} | \mat{\Psi}_{0}, \nu_{0}) \\
	& \propto
	\exp \left\{
	\begin{bmatrix}
		\mat{0}                          \\[0.5em]
		-\frac{1}{2} \mat{K}_{0}         \\[0.5em]
		-\frac{1}{2} \mat{\Psi}_{0}^{-1} \\[0.5em]
		\frac{1}{2} (\nu_{0} - m - 1 + d)
	\end{bmatrix}
	\cdot
	\begin{bmatrix}
		\mat{V} \mat{A}                \\[0.5em]
		\mat{A}^{\top} \mat{V} \mat{A} \\[0.5em]
		\mat{V}                        \\[0.5em]
		\log |\mat{V}|
	\end{bmatrix}
	\right\}.
\end{align*}
The posterior $q(\vec{\mu}, \mat{\Lambda})$ is matrix-normal-Wishart
\begin{align*}
	q(\mat{A}, \mat{V}) & = \N(\mat{A} | \mat{M}, \mat{V}, \mat{K}) \W(\mat{V} | \mat{\Psi}, \nu) \\
	& \hspace{-2.5em} \! \propto \!
	\exp \left\{ \!
	\begin{bmatrix}
		\mat{Y} \mat{W} \mat{X}^{\top}                                      \\[0.5em]
		-\frac{1}{2} (\mat{K}_{0} + \mat{X} \mat{W} \mat{X}^{\top})         \\[0.5em]
		-\frac{1}{2} (\mat{\Psi}_{0}^{-1} + \mat{Y} \mat{W} \mat{Y}^{\top}) \\[0.5em]
		\frac{1}{2} (\nu_{0} - m - 1 + d + \sum_{n=1}^{N} w_{n})
	\end{bmatrix}
	\! \cdot \!
	\begin{bmatrix}
		\mat{V} \mat{A}                \\[0.5em]
		\mat{A}^{\top} \mat{V} \mat{A} \\[0.5em]
		\mat{V}                        \\[0.5em]
		\log |\mat{V}|
	\end{bmatrix}
	\! \right\}.
\end{align*}
The mode mapping and precision of a matrix-normal-Wishart are $\hmat{A} = \mat{M}$ and $\hmat{\Lambda} = (\nu - m) \mat{\Psi}$, respectively. The standard posterior parameters are
\begin{gather*}
	\mat{K} = \mat{K}_{0} + \mat{X} \mat{W} \mat{X}^{\top}, \mat{M} = \mat{Y} \mat{W} \mat{X}^{\top} \mat{K}^{-1}, \\
	\nu = \nu_{0} + \sum_{n=1}^{N} w_{n}, \mat{\Psi} = (\mat{\Psi}_{0}^{-1} + \mat{Y} \mat{W} \mat{Y}^{\top} - \mat{M} \, \mat{K} \, \mat{M}^{\top} )^{-1}.
\end{gather*}

\bibliographystyle{IEEEtran}
\bibliography{references.bib}

\newpage

\vfill

\end{document}